\documentclass[twocolumn]{aastex63}

\usepackage{amsmath}
\usepackage[dvipdfmx]{}

\newcommand{\ccm}{\,\mathrm{cm}^{-3}}
\newcommand{\Msun}{\,\ensuremath{\mathrm{M}_\odot}}

\received{June 1, 2019}
\revised{January 10, 2019}
\accepted{\today}
\submitjournal{ApJ}

\shorttitle{Implications of inhomogeneous metal mixing for stellar archaeology}
\shortauthors{Tarumi et al.}
\graphicspath{{./}{figures/}}

\begin{document}

\title[Implications of inhomogeneous metal mixing for stellar archaeology]{Implications of inhomogeneous metal mixing for stellar archaeology}

\author{Yuta Tarumi}
\affil{Department of Physics, School of Science, The University of Tokyo, Bunkyo, Tokyo 113-0033, Japan}
\affil{Department of Astrophysical Sciences, Princeton University, Peyton Hall, Princeton, NJ 08544, USA}

\author{Tilman Hartwig}
\affil{Department of Physics, School of Science, The University of Tokyo, Bunkyo, Tokyo 113-0033, Japan}
\affil{Institute for Physics of Intelligence, School of Science, The University of Tokyo, Bunkyo, Tokyo 113-0033, Japan}
\affil{Kavli IPMU (WPI), UTIAS, The University of Tokyo, Kashiwa, Chiba 277-8583, Japan}

\author{Mattis Magg}
\affil{Universit\"at Heidelberg, Zentrum f\"ur Astronomie, Institut f\"ur Theoretische Astrophysik, D-69120 Heidelberg, Germany}
\affil{International Max Planck Research School for Astronomy and Cosmic Physics at the University of Heidelberg (IMPRS-HD)}








\begin{abstract}

The first supernovae enrich the previously pristine gas with metals, out of which the next generation of stars form. Based on hydrodynamical simulations, we develop a new stochastic model to predict the metallicity of star-forming gas in the first galaxies. On average, in internally enriched galaxies, the metals are well mixed with the pristine gas. However, in externally enriched galaxies, the metals can not easily penetrate into the dense gas, which yields a significant metallicity difference between the star-forming and average gas inside a halo. To study the consequences of this effect, we apply a semi-analytical model to Milky Way-like dark matter merger trees and follow stellar fossils from high redshift until the present day with a novel realistic metal mixing recipe. We calibrate the model to reproduce the metallicity distribution function (MDF) at low metallicities and find that a primordial IMF with 
slope of $\mathrm{d}N/\mathrm{d}M \propto M^{-0.5}$ from $2\Msun$ to $180\Msun$ best reproduces the MDF. Our improved model for inhomogeneous mixing can have a large impact for individual minihalos, but does not significantly influence the modelled MDF at [Fe/H]$\gtrsim -4$ or the best-fitting Pop~III IMF.

\end{abstract}


\section{Introduction}
\label{sec:intro}

Big Bang nucleosynthesis produced only hydrogen, helium and trace amounts of lithium in the first minutes of the Universe. All metals, as astronomers call elements heavier than helium, were produced in stars and distributed into the interstellar medium (ISM) by their violent deaths. Therefore, the first generation of stars (Population~III, or Pop~III) formed from hydrogen and helium only and the average metallicity of the Universe built up cumulatively over time.

Compared to the present day Universe, where metals and dust provide the main cooling channels during protostellar collapse, primordial gas cools less efficiently \citep{omukai05,bovino16}. The lower cooling rates and higher temperatures in a Pop~III star forming region result in a higher Jeans mass and therefore in larger fragment masses \citep{silk83,bromm99,abel00,bromm02,abel02,yoshida03}. Hence, it is generally expected that the first stars are more massive than present-day stars, although this has not yet been confirmed by direct observations \citep[see reviews by][]{glover05,bromm09,greif15}.
The initial mass function (IMF) of the first stars is crucial to understand the buildup of the first galaxies: the Pop~III stars shape the first galaxies with their radiative and chemical feedback, they may provide the seeds for supermassive black holes, and they contribute to reionisation. Their relative contribution to these processes depends on their mass and therefore on the Pop~III IMF. Despite intensive research in the last years, no studies have conclusively derived the IMF of the first stars, only provided indirect constraints.

No metal-free Pop~III survivor has been discovered in the Milky Way (MW). This implies that most primordial stars accreted metal-rich ISM over their lifetime (\citealt{shen16}, however see \citealt{tanaka17}) or have lifetimes that are shorter than the age of the Universe, which limits the mass of Pop~III stars to $\gtrsim 0.8\Msun$ \citep{2007Salvadori,hartwig15,ishiyama16,magg19}. This also means that we rely on indirect observational constraints.

The direct observation of Pop~III-dominated galaxies at high redshift is very challenging, even with next-generation telescopes \citep{zackrisson11,barrow18,dayal18}. Supernova (SN) explosions of the first stars are very rare and upcoming survey will likely only place upper limits on the number density of massive Pop~III stars \citep{hummel12,hartwig18b,rydberg20}. Gravitational waves from the remnants of the first stars may be detectable with the current sensitivity of ground-based detectors \citep{kinugawa14,hartwig16,belczynski17}. However, to discriminate Pop~III remnant black holes from other formation channels requires a statistically sound sample of several hundred detections over the next decades. Recently, the 21cm sky-averaged signal from the EDGES experiment \citep{edges} has triggered discussions on the contribution from the first stars to gas heating at high redshift \citep{barkana18,fialkov18,mirocha18,schauer19,liu19}.

Numerical simulations of Pop~III star formation are another important tool to predict their nature and IMF \citep{stacy10,greif11b,clark11,hirano14,susa19,shrada20}.
However, current estimates can only approximately treat the gas accretion and mergers of protostars until the stars reach stable hydrogen burning, which limits their predictive power.

The most informative approach to study the nature of the first stars is stellar archaeology \citep{beers05,ji15,frebel15}. As the name suggests, one extracts information from stellar fossils; in this case, one connects the chemical signature of metal-poor stars in the MW to the nucleosynthetic yields from the first SNe \citep{salvadori10,milos18}. Various groups have tried to derive the Pop~III IMF by fitting progenitor stars of different masses to observed extremely metal-poor (EMP) stars with different results: \citet{fraser17} find that their sample of $\sim 30$ EMP stars is best reproduced by a Salpeter IMF for Pop~III stars. However, \citet{ishigaki18} show, based on $R>28000$ spectroscopic abundances of $\sim 200$ EMP stars, that their sample is better fitted by a log-normal Pop~III IMF around $\sim 25\Msun$. In both cases, the method can only probe mass ranges of the Pop~III IMF in which we expect SNe to explode, approximately $10-40\Msun$ and $140-260\Msun$ \citep{heger02}. 
We therefore analyse the metallicity distribution function (MDF), which is a results of an interplay of chemical, radiative, and mechanical feedback from the first and second generation of stars and is therefore sensitive to the complete Pop~III IMF \citep{deB17}.

The MDF is the distribution of stellar metallicities, which is observationally traced by [Fe/H]\footnote{Defined as $[\mathrm{A}/\mathrm{B}] = \log_{10}(m_\mathrm{A}/m_\mathrm{B})-\log_{10}(m_{\mathrm{A},\odot}/m_{\mathrm{B},\odot})$, where $m_\mathrm{A}$ and $m_\mathrm{B}$ are the abundances of elements A and B and $m_{\mathrm{A},\odot}$ and $m_{\mathrm{B},\odot}$ are the solar abundances of these elements.}. If we assume that the metallicity of a star is defined by the metallicity of the molecular cloud out of which it forms, then the stellar metallicity depends on one main ingredient: the ratio of metals to hydrogen in the stellar birth clouds. To understand the MDF and to use it as a tool to study the first stars, we therefore need to understand how metals mix with hydrogen after the first SN explosions.

For simplicity, previous semi-analytical approaches have assumed that metals and hydrogen within the virial radius are homogeneously mixed \citep{deB17,graziani17,visbal18}, that all Pop~II stars have the same metallicity \citep{trenti09,dayal14}, or they applied heuristic models to imitate inhomogeneous mixing \citep{salvadori10,hartwig18a,cote18}. We investigate and improve sub-grid models for metal mixing in this paper, because previous simplistic models limit the reliability of their predictions.

Hydrodynamical simulations have studied the mixing of metals from the first SNe in individual haloes. While \citet{joggerst11} find that the degree of mixing depends on the type of SN, \cite{ritter15} conclude that the abundance of gas clouds is different from progenitor SNe yields due to inhomogeneous mixing. \cite{sluder16} study the evolution of a minihalo after its first Pop~III SN. They conduct cosmological simulation and find that there exists abundance biases, concluding that ``to fully exploit the stellar-archaeological programme of constraining the Pop~III IMF from the observed Pop II abundances, considering these hydrodynamical transport effects is crucial''.

These pioneering studies have highlighted the importance of inhomogeneous metal mixing but our present study is the first attempt to investigate this effect for a cosmological sample of halos. The main scientific question is to understand metal mixing in the first galaxies and how to predict the metallicity of the star-forming gas. These results will help to improve existing semi-analytical models and to deepen our understanding of metal mixing, star formation at high redshift, and the build-up of the MDF.

\section{Methodology: Semi-Analytical Model}

In this section, we summarise the semi-analytical model (SAM) that we apply to simulate the MDF. In the next section, we then present our improved model of metal mixing, derived from 3D cosmological simulations.

The SAM used here has been previously used \citep{hartwig15,hartwig16,hartwig18a,magg16,magg18} and has recently been named \textsc{a-sloth} (Ancient Stars and Local Obervables by Tracing Haloes)\footnote{\url{http://www-utap.phys.s.u-tokyo.ac.jp/~hartwig/A-SLOTH}}. It is based on MW-like dark matter merger trees from the Caterpillar simulation \citep{griffen16}, which represent the hierarchical formation and growth of gravitationally bound structure over time. On top of the dark matter distribution, we model the formation of stars and their chemical and radiative feedback processes. Here, we summarise the main ingredients of \textsc{a-sloth} to reproduce the MDF. More details on the model can be found in previous publications \citep{hartwig15,hartwig18a,magg18}.

\subsection{Pop~III star formation}
The main condition to form Pop~III stars in a minihalo is that molecular hydrogen is the main coolant. Sufficient molecular hydrogen can form if the halo mass is above the critical mass \citep{hummel12,glover13}
\begin{equation}
    M_\mathrm{crit} = 3\times 10^6 \Msun \left( \frac{1+z}{10} \right)^{-3/2},
\end{equation}
which corresponds to a virial temperature of $T_\mathrm{vir}=2200\,\mathrm{K}$.
We do not include the effect of baryonic streaming, which can increase this critical halo mass for gas collapse at high redshift \citep{schauer19b}, but plan to do so in a future study.

Lyman-Werner (LW) photons can photodissociate molecular hydrogen and therefore delay or even prevent the gas collapse in the first haloes. Therefore, we require a second, LW flux-dependent mass threshold for gas collapse \citep{OShea08} and model how the LW background builds up over time based on \citet{greif06}. We also follow the evolution of H{\sc ii} regions around star-forming galaxies and allow star formation in other haloes in these ionised regions only if their virial temperature is $>10^4$\,K.

We allow Pop~III star formation if the gas is sufficiently metal-poor. \cite{2017Chiaki_criterion} argue that we have never observed stars with
\begin{equation}
    10^{\mathrm{[C/H]}-2.30} + 
    10^{\mathrm{[Fe/H]}} \leq 10^{-5.07}.
    \label{eq:transition_criterion}
\end{equation}
and suggest that the truncation is caused by the absence of dust cooling. The C abundance traces the amount of carbon dust, and the Fe abundance traces the amount of Silicate dust. We use Eq.~\ref{eq:transition_criterion} as transition criterion from Pop~III to Pop~II star formation.

If a halo is identified to form Pop~III stars, we define the total stellar mass of this halo as
\begin{equation}
    M_* = \eta_{III} \frac{\Omega _b}{\Omega _m} M_h,
\end{equation}
where the star formation efficiency $\eta_{III}$ needs to be calibrated. This parameter is fully degenerate with baryon fractions of the haloes. For a globally reduced baryon fraction in the haloes, e.g., due to streaming velocities (see \cite{schauer19b}), the same $\eta_\mathrm{III}$ parameter would correspond to a higher physical star formation efficiency. Then, we draw individual stars from a pre-defined IMF (see below) and assign them to this halo until the total mass of stars is $\geq M_*$.

The mass of a Pop~III star defines its lifetime \citep{marigo01,schaerer02} and its final fate \citep{karlsson13}. If the star explodes as a SN, we assume tabulated SN metal yields \citep{kobayashi11,nomoto13} and assume that a fraction, $f_\mathrm{faint}$, of core-collapse SNe explode as faint SNe with the corresponding yields from \citet{ishigaki14}. Based on \citet{ritter15}, we assume that a fraction of the metals, $f_\mathrm{fallback}$, remains in the halo (i.e. falls back after some time) and another fraction, $f_\mathrm{eject} = 1 - f_\mathrm{fallback}$, escapes the gravitational potential of the halo. We assume that these parameters do not depend on the halo mass and we calibrate $f_\mathrm{fallback}$ based on the metallicity distribution function and other constrains such as the external enrichment fraction. For Pop~III stars, we treat these fractions as free parameters, whereas for SNe from Pop~II we calculate the ejected gas mass self-consistently (see below). For the fraction of the metals that escape a Pop~III forming halo, we assume that these metal winds have a constant velocity of $110\,\mathrm{km}\,\mathrm{s}^{-1}$ and may externally enrich neighboring haloes.

After internal enrichment that produces sufficient metals by a Pop~III SN (Eq.~\ref{eq:transition_criterion}), we allow Pop~II star formation after the recovery time $t_\mathrm{rec}$, which we calibrate with priors in the range $10\,\mathrm{Myr} \leq t_\mathrm{rec} \leq 100\,\mathrm{Myr}$ \citep{jeon14,chiaki18}. If a previously pristine halo is sufficiently enriched by external enrichment, we trigger Pop~II star formation in this halo one freefall time after the external enrichment event, which is of the order of $100\,$Myr at the redshifts of interest.

\subsection{Pop~II star formation}
We model the formation of metal-enriched stars with a bathtub model, which will be described in more detail in Magg~et~al. (in prep.). This improved way of simulating Pop~II formation represents a major update to \textsc{a-sloth} as compared to, e.g., \citet{hartwig18a}. The baryonic matter in each metal-enriched, star-forming halo is split into four components:
\begin{itemize}
    \item $M_\mathrm{*,II}$: the mass in metal-enriched stars,
    \item $M_\mathrm{hot}$: the mass of hot gas in the halo,
    \item $M_\mathrm{cold}$: the mass of cold, star-forming gas in the centre of the halo and
    \item $M_\mathrm{out}$: the mass of the outflows, i.e. baryonic gas that has been unbound from the halo.
\end{itemize}
The transition of matter from one component to the other is prescribed according to characteristic time-scales and efficiencies. We aim at keeping the sum of these four components to $(\Omega_\mathrm{b}/\Omega_\mathrm{m}) M_\mathrm{vir}$:
\begin{equation}
    M_\mathrm{*,II}+M_\mathrm{hot}+M_\mathrm{cold}+M_\mathrm{out} = \frac{\Omega_\mathrm{b}}{\Omega_\mathrm{m}} M_\mathrm{vir}.
\end{equation}
We adapt this general stucture from \citet{agarwal12}, i.e., the idea of having a four-component bathtub model, where we keep the the total baryonic mass fraction equal to the cosmic mean \citep[see also][]{cote16}. The exact time-scales according to which the matter moves from one phase to the other and the efficiency of outflows were revised.
Each time a halo forms Pop~II stars the following algorithm is used:
\begin{enumerate}
    \item The halo is initialized. All four baryonic matter components are set to the sum of these components over all progenitors. Thus the mass accretion rate during a time-step of length $\Delta t$ is
    \begin{equation}
    \begin{split}
    \dot{M}_\mathrm{acc} =&  \frac{\Omega_\mathrm{b}}{\Omega_\mathrm{m}\,\Delta t}M_\mathrm{vir}\\-& \left(M_\mathrm{*,II}+M_\mathrm{hot}+M_\mathrm{cold}+M_\mathrm{out}\right)\frac{1}{\Delta t}.\\
    \end{split}
    \end{equation}
    \item Compute relevant time-scales and outflow efficiency:
    \begin{itemize}
        \item $t_\mathrm{ff}$ the free fall time at the 192-fold cosmic over-density, i.e. the density of the halo, for conversion of hot diffuse gas to cold dense gas. It is defined by
        \begin{equation}
        \begin{split}
          t_\mathrm{ff} =& \sqrt{\frac{1}{(1+z)^3\,192\,G\,\rho_\mathrm{m}}}\\ =& 5.2\,\mathrm{Gyr} \left(1+z \right)^{-\frac{3}{2}},
        \end{split}
         \end{equation}
         where $\rho_\mathrm{m}$ is the cosmic matter density at z=0. At redshifts and halo masses of interest, this time-scale is longer than e.g. the cooling time of the gas and thus the gas collapses essentially in free-fall.
        \item $t_\mathrm{dyn}$: the dynamical time-scale of the halo centre, for star-formation from cold gas. We assume that cold gas and stars are concentrated in the central 5 per cent of the virial radius $R_\mathrm{vir}$ of the halo \citep{mo98}. The dynamical time-scale can then be computed as ratio of the virial radius and the circular velocity of this region i.e.
        \begin{equation} 
        \begin{split}
            t_\mathrm{dyn} =& \frac{0.05\, R_\mathrm{vir}}{v_\mathrm{dyn}}\\ =& \sqrt{\frac{\left(0.05\, R_\mathrm{vir}\right)^3}{G\,\left(M_\mathrm{cold}+M_\mathrm{*,II}\right)}},\\
        \end{split}
        \end{equation}
        where the dynamical velocity in the centre of the halo is
        \begin{equation}
            v_\mathrm{dyn} =\sqrt{\frac{G\,\left(M_\mathrm{cold}+M_\mathrm{*,II}\right)}{\left(0.05\, R_\mathrm{vir}\right)^2}}.
        \end{equation}
        \item $\gamma_\mathrm{out}$ the outflow efficiency, which is computed from the energy balance between the injected energy per solar mass of star formation and the needed kinetic energy for outflows (per solar mass of out-flowing mass). This effective specific binding energy is
        \begin{equation}
              \frac{1}{2} v_\mathrm{eff}^2 = \frac{1}{2}\left(v_\mathrm{dyn}^2+v_\mathrm{circ}^2\right).
          \end{equation}
          The circular velocity $v_\mathrm{circ}$ of the halo is defined analogously to the dynamical velocity of the halo
          \begin{equation}
              v_\mathrm{circ} =\sqrt{\frac{G\,M_\mathrm{vir}}{R_\mathrm{vir}^2}}.
          \end{equation}
          For the specific energy input of SNe we find
        \begin{equation}
            e_\mathrm{SN} = 8.7\times 10^{15}\,\mathrm{erg}\,\mathrm{g}^{-1},
        \end{equation}
        which corresponds to one $10^{51}\,\mathrm{erg}$ SN per $57\,\Msun$ of stars formed. For ionizing radiation we find 
        \begin{equation}
            e_\mathrm{ion}= 5.2\times 10^{16}\,\mathrm{erg}\,\mathrm{g}^{-1},
        \end{equation}
          being the equivalent of 30000 ionizing photons per stellar baryon \citep{greif06},  2.0\,eV thermal energy injection for each of these photons and an escape fraction of 10 per cent. We assume that feedback by ionizing photons is only efficient in haloes that are not too far above the atomic cooling limit and smoothly join the two regimes together:
        \begin{equation}
            \gamma_\mathrm{out} = \begin{cases}
                2\frac{e_\mathrm{SN}}{16 \left(v_\mathrm{eff} - 20\,\mathrm{km}\,\mathrm{s}^{-1}\right)^2}  \mathrm{if}\  v_\mathrm{eff}>20\,\mathrm{km}\,\mathrm{s}^{-1}\\
                2\frac{e_\mathrm{SN}+e_\mathrm{ion}}{\left(10\,\mathrm{km}\,\mathrm{s}^{-1}\right)^2}\  \mathrm{else}.\\
            \end{cases}
        \end{equation}
        This assumes that outflows move at 10\,km\,s$^{-1}$ if they are primarily ionization driven and a few times as fast as the escape velocity if they are SN driven. The choice of the exact transition between the two regimes is calibrated to reproduce, e.g. the stellar mass to halo mass relation of the MW (see below). 
    \end{itemize}
    \item We solve the following differential equations with a simple forward Euler method:
    \begin{equation}
        \dot{M}_\mathrm{*,II} = \eta_{II} \frac{M_\mathrm{cold}}{t_\mathrm{dyn}}
    \end{equation}
    \begin{equation}
        \dot{M}_\mathrm{out} = \gamma_\mathrm{out}\,\dot{M}_\mathrm{*,II}
    \end{equation}
    \begin{equation}
        \dot{M}_\mathrm{cold} = - \dot{M}_\mathrm{*,II} -\dot{M}_\mathrm{out} + \frac{M_\mathrm{hot}}{t_\mathrm{ff}}
    \end{equation}
    \begin{equation}
        \dot{M}_\mathrm{hot} =  -\frac{M_\mathrm{hot}}{t_\mathrm{ff}} + \dot{M}_\mathrm{acc},
    \end{equation}
    where the Pop~II star formation efficiency $\eta_{II}=0.1$ was calibrated to reproduce the stellar mass to halo mass relation (see below).
    \item Afterwards we update the produced metals and feedback:
    \begin{itemize}
        \item reduce metal mass because of outflows
        \begin{equation}
             M_\mathrm{metals, t+\Delta t} = M_\mathrm{metals, t} \frac{M_\mathrm{cold}+M_\mathrm{hot}}{M_\mathrm{gas}},
        \end{equation}
        where the total gas mass is 
        \begin{equation}
        \begin{split}
            M_\mathrm{gas}= &M_\mathrm{cold}+M_\mathrm{hot}\\&+\Delta M_\mathrm{*,II, tot}+ \Delta M_\mathrm{out,tot},\\
        \end{split}
        \end{equation}
        and $\Delta M_\mathrm{*,II, tot}$ and $\Delta M_\mathrm{out,tot}$ are the mass of stars and outflows formed during the whole time-step.
        \item Add metals generated during this time-step
        \begin{equation}
            M_\mathrm{metals, t+\Delta t} = M_\mathrm{metals, t} + 0.05 \Delta M_\mathrm{*,II, tot} ,
        \end{equation}
        where we took the IMF averaged metal-yields of 0.05\Msun\ per 1\Msun\ of star formation from \citet{vincenzo16}.
        \item We use snowplough algorithm to compute the size of the metal-enriched region and propagate the ionization front as described in \citet{magg18}.
    \end{itemize}
    \end{enumerate}
This new model comes with several free parameters. We calibrate these parameters mainly by the stellar mass to halo mass relation in a mass range of 2 dex. For more details see Sec.~\ref{sec:results}.

\section{Methodology: Metal Mixing}
    
    In this section, we explain our approach to investigate inhomogeneity of metallicity in the high-redshift ISM. Although its importance has been pointed out in previous research \citep{cote18}, metallicity inhomogeneity has often been neglected in archaeological approaches. However, without understanding this effect we cannot properly connect the observational information on EMP stars and theoretical efforts on SN yields. We take into account the inhomogeneity by post-processing the metallicity of each star-formation event in {\sc a-sloth}, according to the distribution extracted from cosmological simulations.
    \subsection{Cosmological Simulation}

    To study metal mixing in high-redshift galaxies, we analyse the Renaissance simulations \citep{xu16}. The simulations were conducted with the adaptive mesh refinement (AMR) code \textsc{enzo} \citep{bryan14}. The boxsize is 28.4 cMpc/h, the particle masses of dark matter particles are $2.9\times 10^{4} \Msun$, and the spacial resolution is $19$ comoving pc at the most resolved region. This corresponds to a physical resolution of a few parsec at the redshifts of interest. 
    The Renaissance simulations are suited for our purpose because they have a specific treatment for Pop~III stars (including IMF), and the mass resolution is sufficient to resolve $3 \times 10^{6} \Msun$ haloes, where the formation of Pop~III stars takes place \citep{xu16}. In less massive haloes, Pop~III star formation may be delayed or suppressed by baryonic streaming velocities \citep{tseliakhovich10,schauer19b}.

    For more information, the simulations are well described in \citet{Oshea15} and \citet{xu16}. The simulations consist of three distinct boxes, ``Rarepeak'', ``Normal'', and ``Void'', with the names representing the overdensity on super-halo scale. Since we want to study the properties of galaxies in a cosmologically representative environment, we select the ``Normal'' region as the main analysis set. At $z=12$, 2223 haloes contain Pop~III stars.
    
    \subsection{Sample selection}
    We select halos to analyse based on two criteria: First, we select haloes that contain gas denser than $1\ccm$ in order to compare the metallicity between dense gas and average gas within a halo. In the following, we define ``dense gas'' as gas denser than $1\ccm$. We chose $1\ccm$ to improve statistics because not many of the simulated galaxies have gas denser than 10 (1122 haloes) or 100$\ccm$ (433 haloes), whereas 2733 haloes contain gas cells denser than 1$\ccm$. If we only considered cells that form stars in the next timestep, only six halos would contain more than 10 cells. For all the haloes with gas above a density of 100$\ccm$ we verified that the metallicities above all three density thresholds are similar.

    Second, to make sure that the halo is well resolved, we apply two conditions: haloes which include more than 1000 cells in its virial radius, and haloes in which the dense gas is resolved by at least 10 resolution elements. The numbers of analysed haloes are 2733, 6150, 1145 for Normal z=12, Rarepeak z=15, and Normal z=15.
    
    \subsection{Metallicity Shift dZ}
    
\begin{figure}
	\includegraphics[width=\columnwidth]{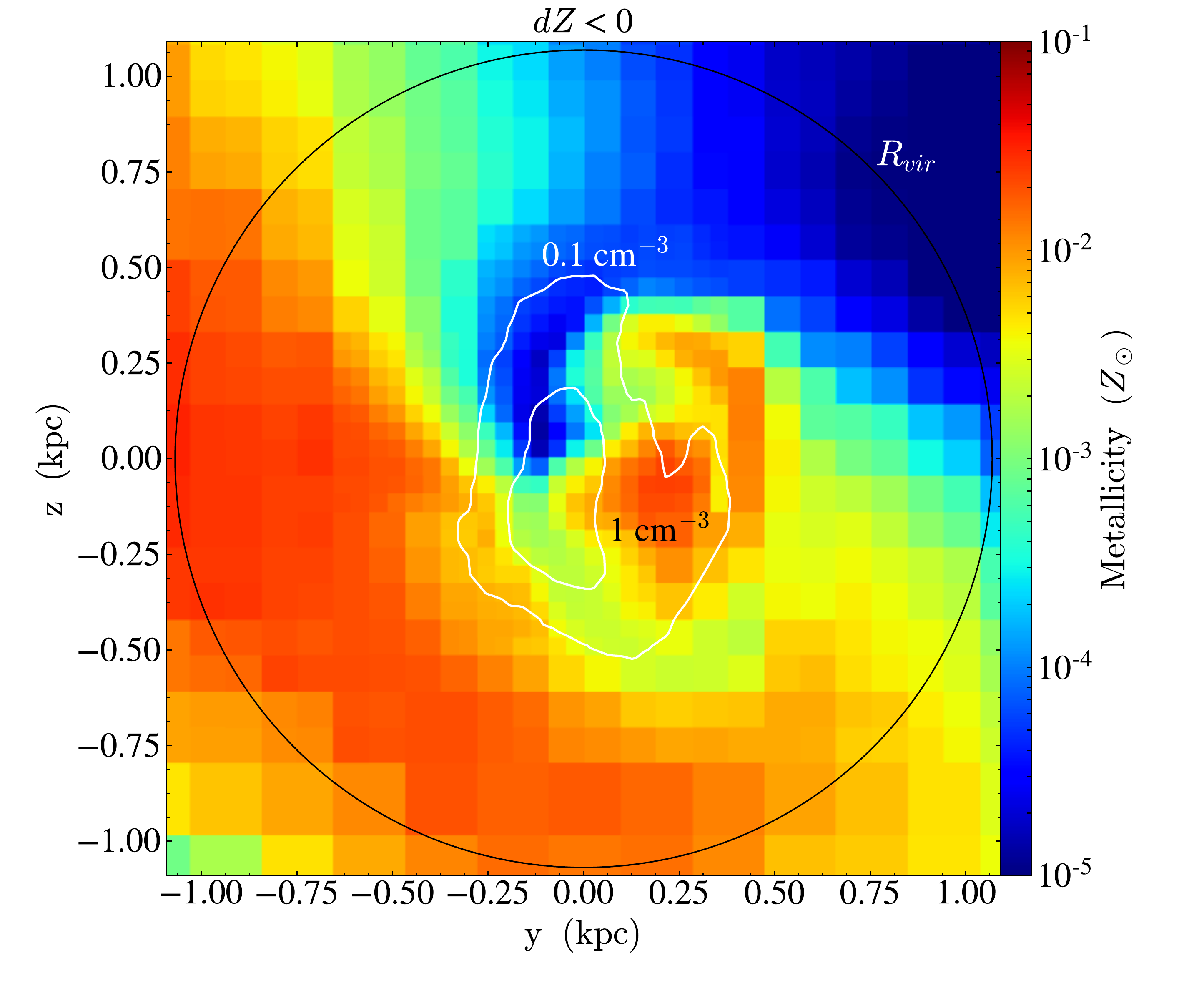}
	\includegraphics[width=\columnwidth]{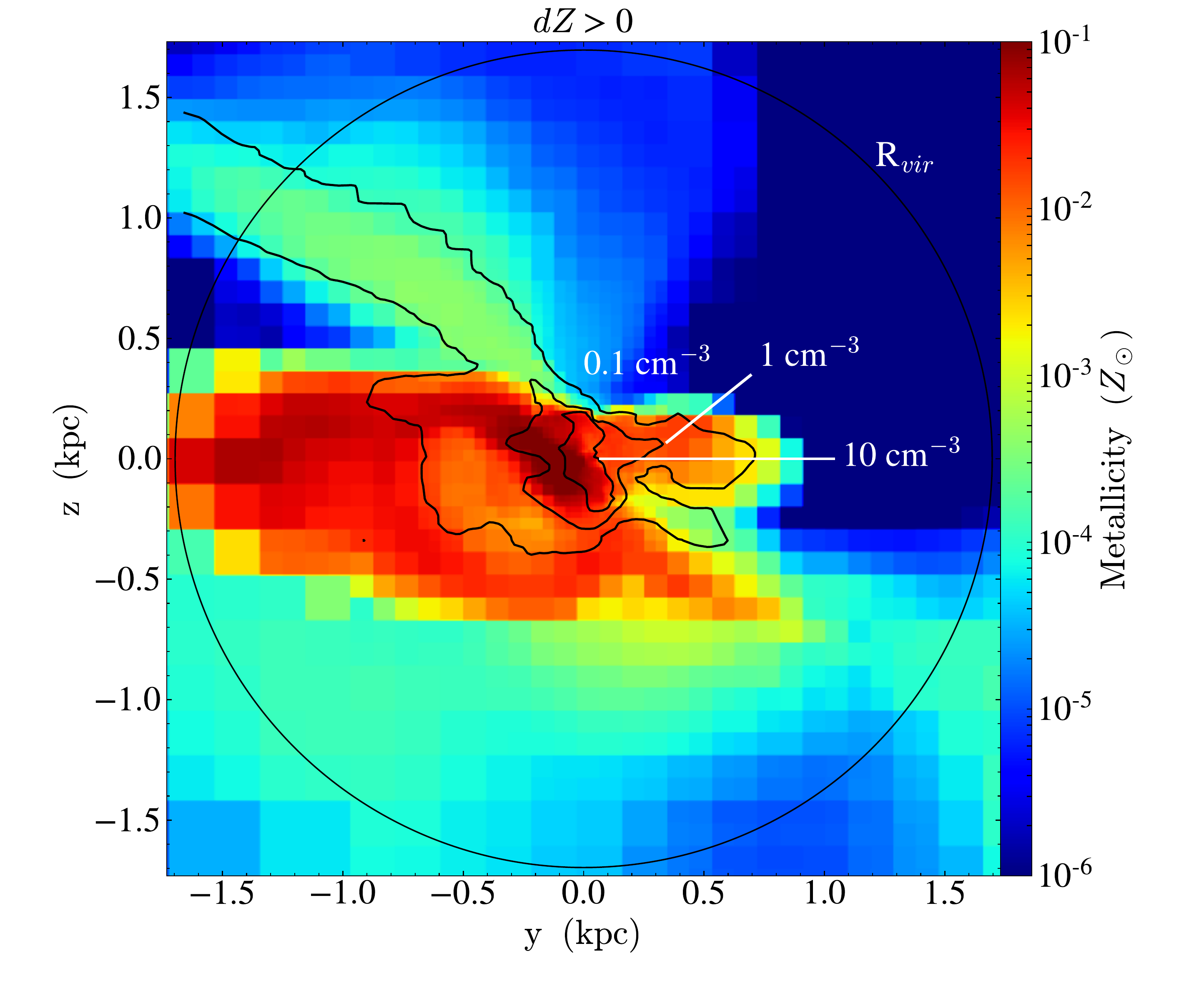}
    \caption{Slice plots 
    showing metallicity distribution of haloes with dense gas at redshift $z=12$. In both panels, the large black circle represents the virial radius of the halo, and the white/black curves are iso-density contours. The top panel illustrates a halo that has negative $dZ$ ($dZ$ = -0.28), suggesting that gas collapse occurs before metal enrichment, and metals can not penetrate into the dense gas. The bottom panel illustrates a halo that has positive $dZ$ ($dZ$ = 0.14), suggesting that gas collapse occurs after metal enrichment. The masses of the dark matter haloes presented in top and bottom panel are $7.4\times 10^{7} \Msun$ and $3.0\times 10^{8} \Msun$, respectively.}
    \label{fig:Sliceplot}
    \end{figure}

    To predict the metallicity of star-forming gas we introduce the ``metallicity shift'' $dZ$ as
    \begin{equation}
        dZ=Z_\mathrm{dense}-Z_\mathrm{all},
    \label{eq:dZ}
    \end{equation}
    where $dZ$ quantifies the difference of metallicity $Z$\footnote{Defined as $Z = \log_{10}(m_\mathrm{metal}/m_\mathrm{H})-\log_{10}(m_{\mathrm{metal},\odot}/m_{\mathrm{H},\odot})$, where $m_\mathrm{metal}$ is the abundance of metals, $m_\mathrm{H}$ is the abundance of Hydrogen, and $m_{\mathrm{metal},\odot}$ and $m_{\mathrm{H},\odot}$ are the solar abundances of these \citep{asplund09}.} between dense gas and all gas inside each halo. Positive $dZ$ represents a situation where dense gas is metal-rich compared to the average gas of the halo. In Fig.~\ref{fig:Sliceplot} we show two exemplary slice plots showing negative $dZ$ and positive $dZ$ haloes. The mean gas metallicity, $Z_\mathrm{all}$, is available in many models with coarse resolution. Together with an analytical prediction of $dZ$, this allows to calculate the actual metallicity of star-forming gas.
    
    \subsubsection{Physical interpretation}
    Various processes are at play to provide positive or negative values of $dZ$. The positive shifting process can be related to metal line cooling. Since metal-rich gas cools efficiently by metal and dust cooling, it is likely to be dense. There are two processes with the opposite effect. One process is shielding. Metal-rich winds cannot penetrate dense gas clumps easily \citep{jeon14,chen17b,chiaki18}. If gas cooling and clump formation occur earlier than metal enrichment, the metallicity shift may take a negative value. The other process is the feedback origin. Metal-rich gas tends naturally to be hot, because thermal and chemical feedback sources are the same \citep{emerick18}. 
    The interplay of these processes, together with inhomogeneous mixing of metals with pristine gas manifest in the distribution of $dZ$.
    
    \subsubsection{Correlation analysis}
    The goal is to find an analytical expression to predict $dZ$ for a given halo. We expect $dZ$ to be correlated to other halo properties. For example, stellar mean age traces the time that has passed after a star formation event. Galaxies with older stellar populations have longer time to mix after energy injection. If we wait longer than the cooling time of metal-rich gas, one could expect that $dZ$ takes a positive value by the first process (and disappearance of the third process) explained above.
    
\begin{figure}
	\includegraphics[width=\columnwidth]{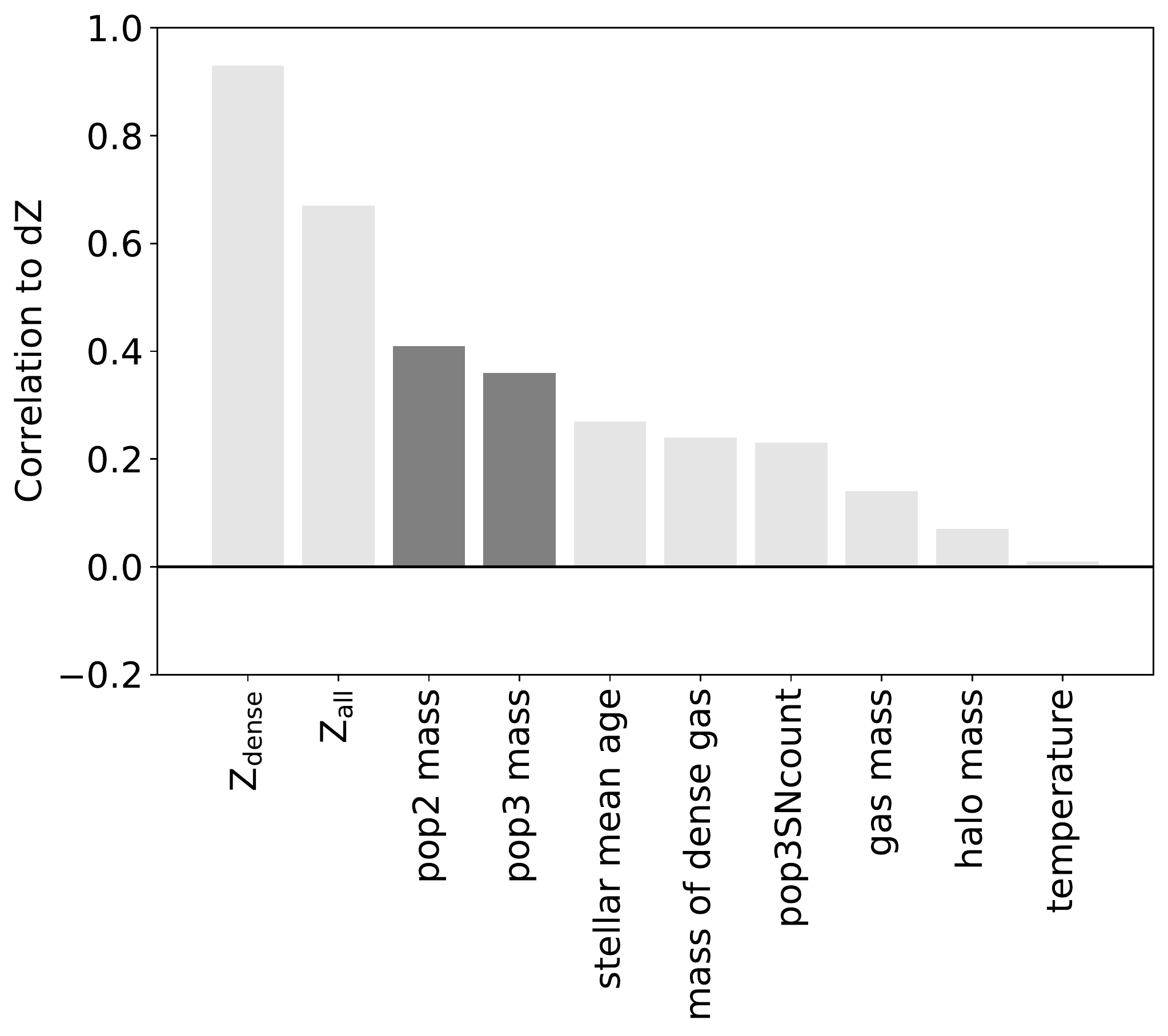}
    \caption{Pearson's correlation coefficients between $dZ$ (the metallicity shift defined in Eq.~(\ref{eq:dZ})) and other halo properties. 
    $dZ$ has strong correlation to metallicity of dense gas and metallicity of all gas, which is natural because they are defined with these quantities. Also $dZ$ has mild correlation to Pop~II stellar mass and Pop~III progenitor stellar mass. These correlations come from the bimodality of $dZ$ distinguished by whether haloes contain stars or not. 
    }
    \label{fig:non_correlation}
    \end{figure}

\begin{figure}
	\includegraphics[width=\columnwidth]{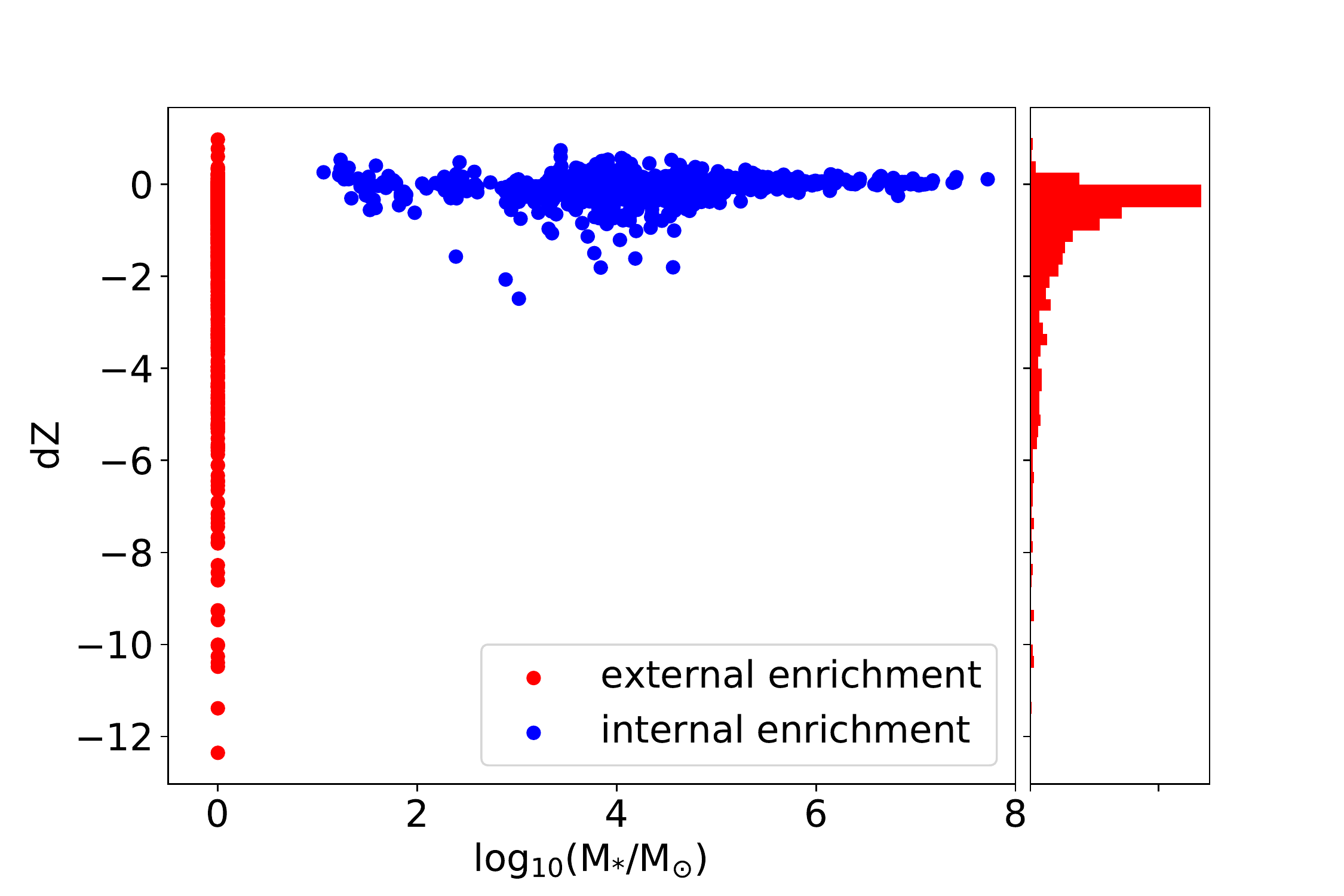}
    \caption{Scatter plot between $dZ$ and mass of all stars in a halo. This figure clearly shows a bimodality, suggesting that internal enrichment and external enrichment behave very differently in terms of $dZ$. We labeled haloes without stars as ``external enrichment'' because they are dominated by external enrichment, and haloes with stars as ``internal enrichment'' because they are dominated by internal enrichment. Since the horizontal axis is logarithmic, we artificially set 1 $\Msun$ for haloes without any stars for illustration purpose. On the right we show the histogram of $dZ$ for externally enriched haloes.
    }
    \label{fig:dZ_Mstars_bimodality}
    \end{figure}

    We calculate Pearson correlation coefficients between halo properties and the metallicity shift. We include the number of Pop~III SNe (which traces injected energy by Pop~III, pop3SNcount), halo mass (halo mass), Pop~III mass that went into SNe (pop3 mass), Pop~II stellar mass (which traces injected energy, pop2 mass), gas mass (gas mass), metallicity of all gas (Z$_\mathrm{all}$), mean temperature (temperature), stellar mean age (which traces the time passed after star formation event, stellar mean age), metallicity of dense gas (Z$_\mathrm{dense}$), and mass of dense gas (mass of dense gas). First, we calculate the correlation matrix among various quantities in search of explanatory variables for $dZ$. Pearson's correlation coefficients are presented in Fig.~\ref{fig:non_correlation}. The two strongest correlations to $dZ$ are dense gas metallicity and average gas metallicity, which is natural because $dZ$ is defined based on these two quantities. The absence of any other strong correlations demonstrates that metal mixing in the first galaxies is an intrinsically stochastic process. The mass of Pop~II stars shows the next strongest correlation. When we take a close look at the actual scatter plot between $dZ$ and mass of all
    stars, two distinct clusters are observed (Fig.~\ref{fig:dZ_Mstars_bimodality}). $dZ$ behaves very differently between haloes with stars and those without stars.
    This bimodality also explains other correlations to $dZ$, namely number of Pop~III SNe, and Pop~III mass that went into SNe. If a halo has a finite metallicity but did not experience SNe before, it must have been enriched externally by at least one SN from a nearby halo.
    
    We do not see any other quantities that are correlated to $dZ$. In particular, we do not find a dependence of the metallicity shift on redshift or halo masses.
    
    \begin{figure}
	\includegraphics[width=\columnwidth]{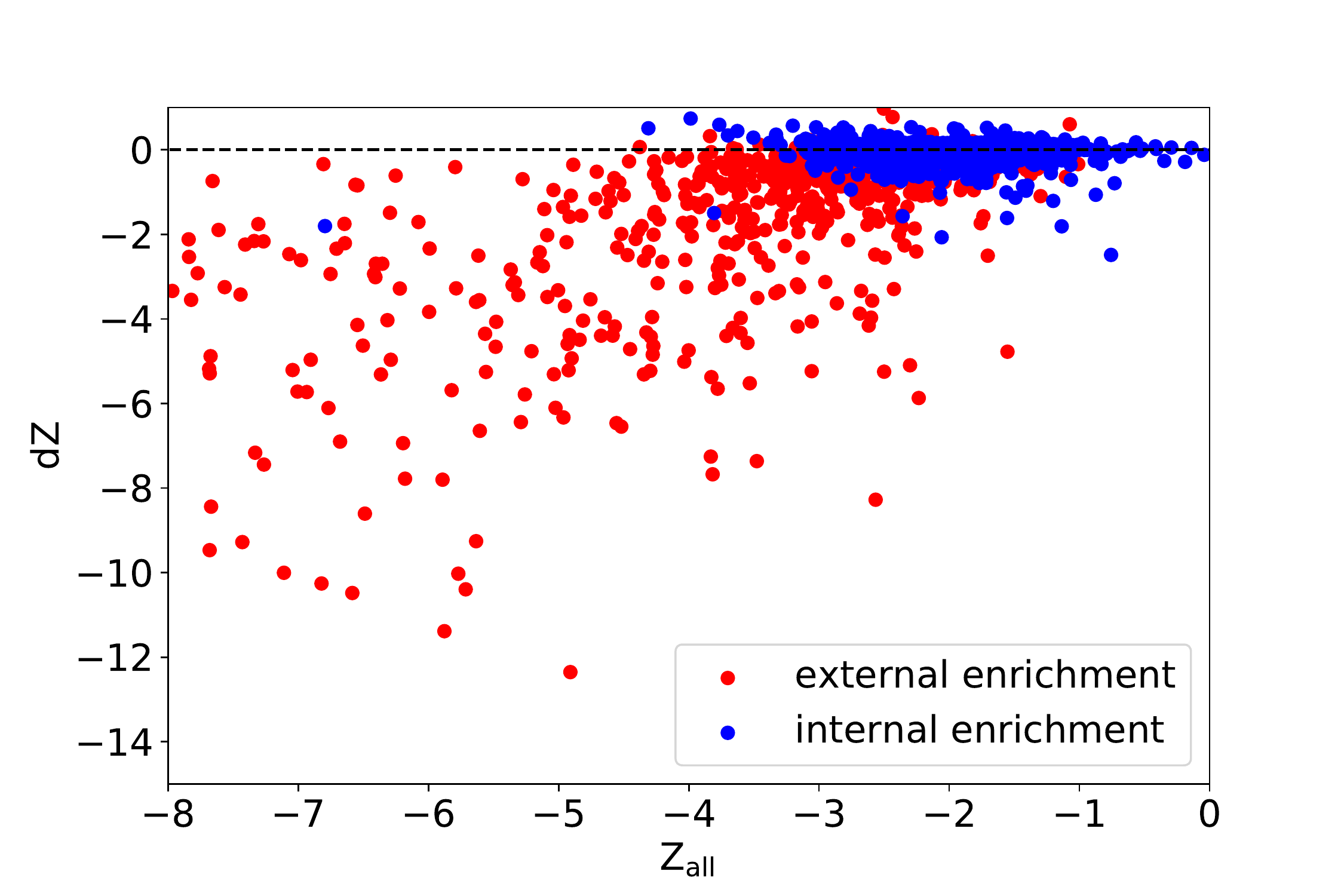}
    \caption{
    Scatter plot between average metallicity of all gas and metallicity difference between dense gas and average gas, $dZ$. The trend suggests that gas at [Fe/H]$ \gtrsim -4$ is dominated by internal enrichment, where the SN energy can efficiently mix all gas components, and metal-poor gas is dominated by external enrichment. [Fe/H] $\simeq -4$ corresponds to the critical metallicity for Pop~II star formation in the Renaissance simulation.}
    \label{fig:Z_dZ_overall}
    \end{figure}

    \subsubsection{Overall Trend}
    Fig.~\ref{fig:Z_dZ_overall} shows a scatter plot between average metallicity inside each halo and $dZ$. Internally enriched haloes reside in relatively high-metallicity region, while externally enriched haloes are widely scattered on the figure. In internal enrichment, where the direct energy injection by SNe strongly disrupts the host halo and creates turbulence, metals produced by SNe mix with the surrounding gas efficiently. It is expected that in such haloes $dZ$ is close to zero. On the other hand, in external enrichment, where produced metals are ``just'' accreted onto the gas cloud, mixing is not efficient. In a case where gas collapses earlier than external enrichment, $dZ$ is expected to be negative, because metals cannot penetrate into the gas clouds that are already dense. Since internal enrichment pollutes each halo to higher metallicity compared to external enrichment, it is natural to interpret the figure that the increasing trend in $dZ$ represents the transition of enrichment mode from external enrichment to internal enrichment.
    
    \subsection{Internal Enrichment}
    
    We identify internally enriched haloes with haloes that contain at least one star that went into SN. In internally enriched haloes, almost no correlation is observed between metallicity and $dZ$, therefore we regard them as metallicity-independent and we fit the distribution with a Gaussian distribution function with $\mu = -0.03, \sigma = 0.15$:
    \begin{equation}
        p(x) = \frac{1}{\sqrt{2\pi (0.15)^{2}}}\exp\biggl[\frac{(x+0.03)^2}{2 (0.15)^{2}}\biggr]
        \label{eq:Gaussian}
    \end{equation}
    The mean $dZ$ is almost zero, however slightly negative, suggesting that on average star-forming gas has almost the same metallicity as the average gas of the halo. 
    
    However, the distribution is a bit skewed, with a longer tail on the negative end. This can also be seen by calculating the mean and standard deviation directly from the data, instead of fitting a Gaussian distribution. The mean and standard deviation of the data points is -0.08 and 0.27\,dex.
    
    \subsection{External Enrichment}
    We identify externally enriched haloes with haloes that do not contain any stars that went into SN. The importance of such external enrichment for the formation of EMP stars has already been pointed out by \citet{smith15}, although \citep{jaacks18} show that external enrichment alone may not be sufficient to reach the critical metallicity (Eq. \ref{eq:transition_criterion}) to trigger Pop~II star formation. In externally enriched haloes, an obvious increasing trend is observed between metallicity and $dZ$. We therefore bin the metallicity with $\Delta Z = 1$ in the range $-5 < Z \leq -1$, and in $Z \leq -5$ we group them together. We calculate mean and standard deviation in each bin. We fit the distribution functions of $dZ$ at different metallicities with an exponentially modified Gaussian distributions. The distribution has three free parameters, $(K, \mu, \sigma)$, and the probability distribution function $p(x; \mu, \sigma, \lambda)$ is
    \begin{equation}
        p(x; \mu, \sigma, \lambda)=
\frac{\lambda}{2} \exp\biggl[\frac{\lambda}{2}(2\mu+\lambda\sigma^{2}-2x)\biggr]{\rm erfc}\biggl(\frac{\mu+\lambda\sigma^2-x}{\sqrt{2}\sigma}\biggr), \\
    \label{eq:PdZ_external}
    \end{equation}
    where
    \begin{eqnarray}
        {\rm erfc}(x) 
        &=& 1 - {\rm erf}(x)\\
        &=& \frac{2}{\sqrt{\pi}}\int^{\infty}_{x}e^{-t^{2}}dt.
    \label{eq:erfc}
    \end{eqnarray}
    The fitting results, which can be used to implement a sub-grid model for incomplete mixing, are presented in Table.~\ref{tab:fitting_result}. The table shows the evolution of the distribution function of $dZ$ with mean metallicity.
    
    \begin{table}
	\centering
	\caption{Best-fitting parameters for external enrichment. These parameter sets are for ``$-dZ$'', not $dZ$ itself.} 
	\label{tab:fitting_result}
	\begin{tabular}{l|lll} 
		\hline
		Metallicity & $\mu$ & $\lambda$ & $\sigma$ \\
		\hline
		$-2<Z\leq -1$ & -0.07 & 0.14 & 2.78\\
		$-3<Z\leq -2$ & -0.01 & 0.24 & 1.41\\
		$-4<Z\leq -3$ & 0.08 & 0.20 & 0.77\\
		$-5<Z\leq -4$ & 0.16 & 0.19 & 0.38\\
		$-8<Z\leq -5$ & 1.63 & 1.01 & 0.35\\
	\end{tabular}
\end{table}

    \subsection{Implementation}
    We implement this new recipe for improved metal mixing in \textsc{a-sloth}, which constitutes a major improvement compared to earlier versions of the code. We discriminate the internally enriched and externally enriched haloes by the stellar mass inside haloes. If a halo has already experienced star formation (either Pop~II or Pop~III), we apply the internal enrichment formula. Otherwise, we apply the external enrichment formula, equation~(\ref{eq:PdZ_external}), based on pre-calculated look-up tables.
    
    \subsection{Comparison to other research}
    
    One of the limitations of our metal-mixing model is the finite resolution of the Renaissance simulations. These may not allow us to capture inhomogeneities at the highest densities. In the high-resolution simulations of \cite{greif10} the metallicity of the recollapsing halo becomes uniform, implying almost zero metallicity shift for internally enriched haloes. \cite{chiaki18} show several cases of internal enrichment where the metallicity of the recollapsing gas is much lower than the average metallicty of the halo, implying a negative metallicity shift that is potentially larger than what we find. Both simulations, as well as \cite{chen17b} and \cite{smith15} show that the densest parts of externally enriched haloes are usually not enriched efficiently. This is consistent with the strongly negative metallicity shift we find for externally enriched haloes. However, all simulations of this type focus on one or a few haloes. The low-number statistics of the high-resolution simulations do not yet allow a statistical description of the inherently chaotic process of metal mixing.
    
\begin{figure}
	\includegraphics[width=\columnwidth]{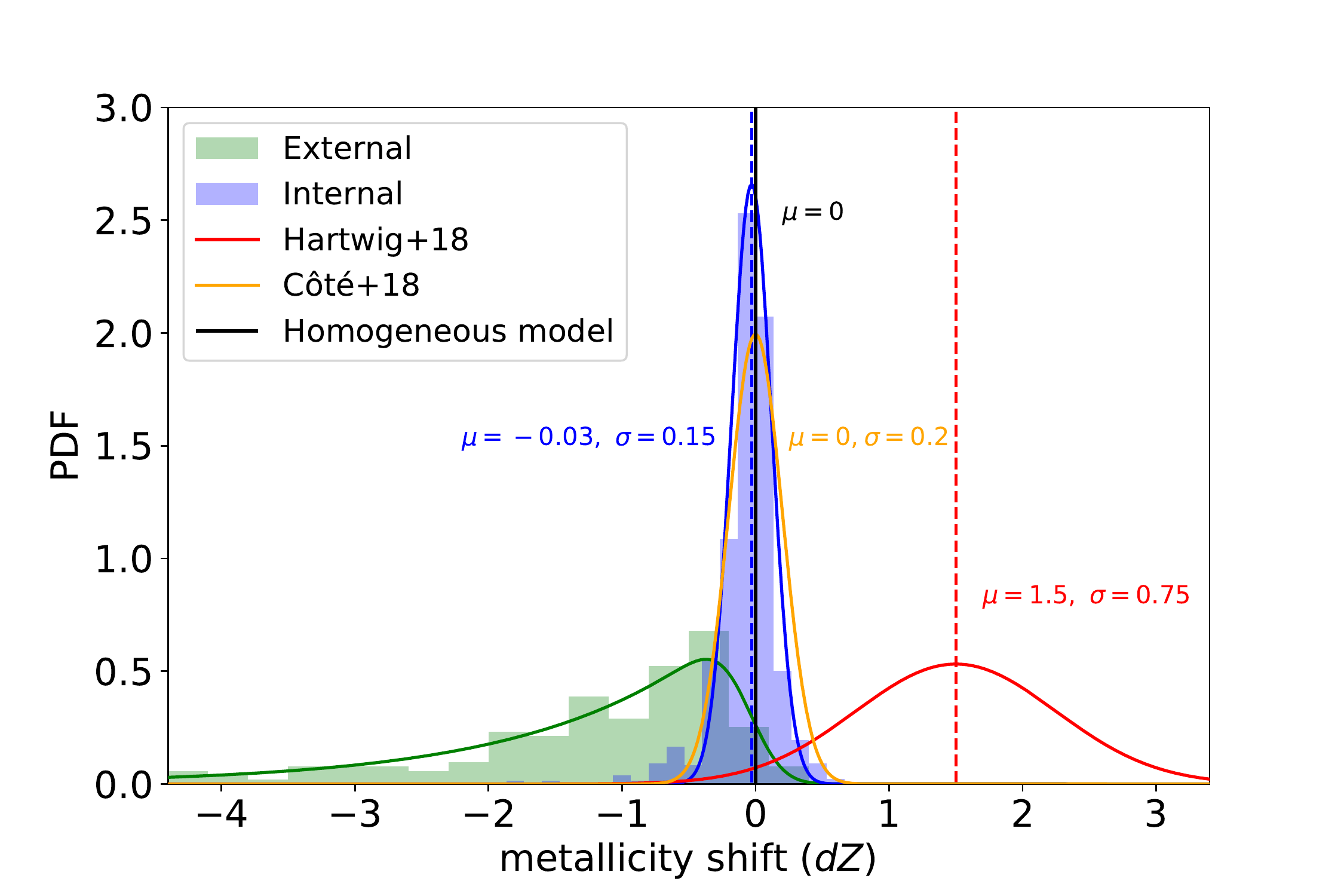}
    \caption{Comparison of our new metallicity shift treatment (green, blue) to previous implementations. Our sample is taken from the Normal $z=12$ dataset (histograms). The green curve is the fitted distribution function of $dZ$ among externally enriched haloes, exemplarily for the metallicity range $-4<\mathrm{Z}\leq -3$. The blue curve is the fitted distribution function of $dZ$ among internally enriched haloes. For comparison, we also show the $dZ$ distribution of a 
    SAM considering inhomogeneous metallicity inside each galaxies \citep{cote18} and of a homogeneous model \citep[$\delta$-function at zero, e.g. used in][]{deB17}. Our mean metallicity shift is negative for both internal and external enrichment, and we see a prominent long tail towards the negative end in the external enrichment.
    }
    \label{fig:metal_mixing_distribution}
    \end{figure}

    In Fig.~\ref{fig:metal_mixing_distribution} we compare our new implementation of the metallicity shift to previous approaches. In one SAM \citep{deB17} this metallicity shift was not taken into consideration, and their treatment corresponds to $dZ = 0$ for all haloes. In another SAM \citep{cote18} metallicity inhomogeneity is taken into account by convolving a Gaussian with $\mu = 0,\ \sigma = 0.2$ with the final MDF. 
    The authors report that they need the convolution to reproduce the MDF extracted from hydrodynamical simulation with their SAM. In our previous implementation, we assumed that produced metals do not mix with all the hydrogen and therefore the metallicity shifts were positive values \citep{hartwig18a}. The cosmological simulation, however, suggests that the mean metallicity shift is negative for both internal and external enrichment. Previous authors did not try to derive such a metallicity shift. However, their approaches and implementations can be interpreted in our new framework as metallicity shifts. In this comparison, the implementation of \citet{cote18} matches very well our method for internal enrichment.
    
    \citet{sarmento17,sarmento19} and  \citet{safarzadeh18} use elaborate sub-grid model to keep track of the pristine gas fraction in each cell. Their model let Pop~III stars form even in enriched cells. Such treatment is suited to follow ``residual'' Pop~III star formation in enriched regions. However, the main purpose of our $dZ$ is to predict the metallicity of stars (or their progenitors, star-forming gas), taking inhomogeneity of metallicity into account. For such purpose, direct analysis of cosmological simulation is the best way to extract this information. 
    
    \citet{hirai17} included their sub-grid metal diffusion recipe. The model calculates the amount of metals diffused to the next cells by the metallicity gradient, shear tensor of the cells, and a scaling factor for metal diffusion. They use elemental abundance patterns to calibrate their sub-grid model and conclude that the metal mixing timescale is less than 40 Myr, shorter than the dynamical time of the typical dwarf galaxies. This comparatively short mixing timescale means that gas and metals are well-mixed, which is consistent to our overall trend that the typical $dZ$ is close to zero. A halo with very negative $dZ$ can be produced if the collapse of a gas cloud happens earlier than the mixing timescale after the first SNe. Such haloes also exist in the simulation, see e.g. the top panel of Fig.~\ref{fig:Sliceplot}.
    
    An alternative approach is to describe metal mixing as a diffusion process \citep{karlsson08, komiya20}. They assume diffusion coefficients that allows galactic gas to be mixed well within a short period of time ($\simeq 30$ Myr). This is consistent to our finding that in internally enriched haloes gas is mixed quite well.

\section{Results}
\label{sec:results}

    In this section, we will first present the calibration of our model based on the MDF and discuss the effect of metal mixing. Then, we will demonstrate that this calibrated model is also able to reproduce additional, independent observations.
    
    \subsection{MDF}
    To calibrate our theoretically predicted MDF we use a de-biased MDF in the range of $-4 \leq \mathrm{[Fe/H]} \leq -3$ provided by \cite{2020Youakim_MDF}. This MDF is based on the photometric Pristine survey \citep{starkenburg17}, corrected for all major biases. This metallicity range is strongly affected by the properties of Pop~III stars and dominating the statistical comparison. For our model prediction, we exclude stars in simulated satellite galaxies to guarantee that we compare halo stars to halo stars.
    
    To quantify the fit quality, we calculate the Kolmogorov-Smirnov (KS) distance for each MDF from each merger tree, i.e., the maximum distance between the observed and modelled cumulative MDF in the range $-4 \leq \mathrm{[Fe/H]} \leq -3$. We use 30 independent MW-like halo trees from the Caterpillar simulation. First, we execute the model, and obtain the MDF as model prediction on each tree. Next, we calculate the KS distance on each MDF. Finally, we use the average of the 30 KS distances as the quantification of comparison between observation and model prediction.

    \begin{table}
	\centering
	\caption{Parameter values in our fiducial model. This set of parameters best reproduces the MDF at $-4\leq$ [Fe/H] $\leq-3$ as we show below. We fixed the Pop~III metal fallback fraction at $f_\mathrm{fallback} = 1 - f_\mathrm{eject}$.}
	\label{tab:fiducial parameter}
	\begin{tabular}{ll} 
		\hline
		Parameter & Value\\
		\hline
		Pop~III SFE & $\eta_{\rm{III}} = 1\times 10^{-2}$\\
		Pop~III metal ejection fraction & $f_{\rm{eject}} = 80\%$\\
		lower IMF limit & $M_{\rm{min}} = 2 {\rm M}_{\odot}$\\
		upper IMF limit & $M_{\rm{max}} = 180{\rm M}_{\odot}$\\
		IMF slope & $\alpha = 0.5$\\
		recovery time & $t_\mathrm{recov} = 30$ Myr
	\end{tabular}
\end{table}

     We calibrate the model parameters by minimizing the average of KS distances to the observed de-biased MDF \citep{2020Youakim_MDF}. We present the parameters of our fiducial model in Table~\ref{tab:fiducial parameter}. We find a top-heavy IMF with the slope $\alpha = 0.5$ in the mass range from $2\Msun$ to $180\Msun$, with a Pop~III SFE of $1\%$ to best reproduce the MDF. For this set of parameters, the average of KS distance is 0.074. We also estimate p-value from KS distances assuming an observational sample size of 2762, based on the sum of the ``corrected''-row in Table.A1 of \cite{2020Youakim_MDF}. The average p-value over 30 merger trees is 0.018. It means our calibrated MDF and observation are in some tension, but the difference is not statistically significant at 99\% significance level. This tension is partly due to the fact that not all merger trees are equally representative of the merger history of the MW. The highest p-value for one MW-like merger tree with the fiducial parameters is 0.49, showing the importance of variations in the merger history.
     
\begin{figure}
	\includegraphics[width=\columnwidth]{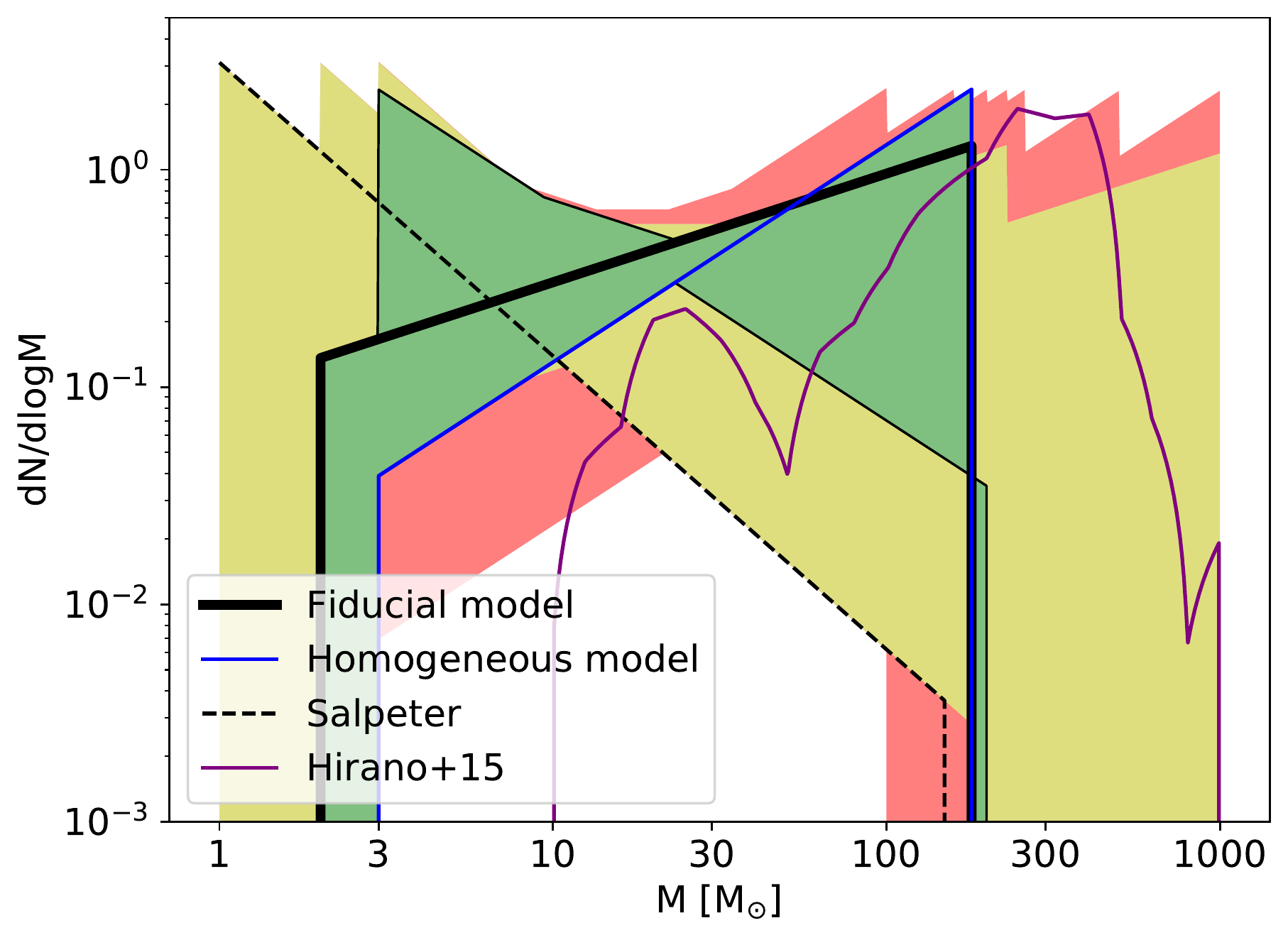}
    \caption{Comparison of the predicted primordial IMF. The black line illustrates our fiducial model which minimizes the average KS distance for 30 trees. The blue line illustrates the calibration assuming homogeneous metal mixing. The green/yellow shaded region is the IMF range that has average p-value more than 90\%/10\% of the one obtained with fiducial parameter. The red shaded region includes all tested IMFs. The dashed line represents the Salpeter slope to guide the eye. With red line we also overplot the Pop~III IMF by \citet{hirano15} derived from numerical simulations. Taking inhomogeneity into account slightly modifies the IMF prediction but the effect is within the uncertainty of the model.}
    
    \label{fig:IMF comparison}
    \end{figure}

     In Fig.~\ref{fig:IMF comparison} we compare our derived fiducial IMF and an independent IMF obtained from numerical simulations \citep{hirano15}. The green region shows the range of best-fitting IMFs, i.e., p-value more than 90\% of the fiducial model. The yellow region illustrates the marginally well-fitting IMFs, i.e., p-value more than 10\% of the fiducial model. The red region represents the disfavoured IMFs, i.e., p-value less than 10\% of the fiducial model. The homogeneous model is within the best-fitting region, therefore it is statistically indistinguishable from our best-fitting model with inhomogeneity. All the best-fitting IMFs are more top-heavy than the Salpeter IMF. Our IMF favours stars with $2 \sim 200 \Msun$. A large fraction of PISNe from Pop~III stars was not favoured (see also \citet{deB17} and \citet{salvadori19}).
     
     Our calibration is not very sensitive to the lower mass limit of the Pop~III IMF, because such low-mass stars do not contribute to chemical enrichment and therefore do not directly affect the MDF. Also, our calibration based on the KS test is not very sensitive to the low-metallicity tail of the MDF, because of the small number of observed stars in this range. Therefore, our improved model for metal mixing, which mostly affects star formation at very low metallicities, does not affect the IMF calibration significantly.

    \begin{figure}
        \centering
        \includegraphics[width=\columnwidth]{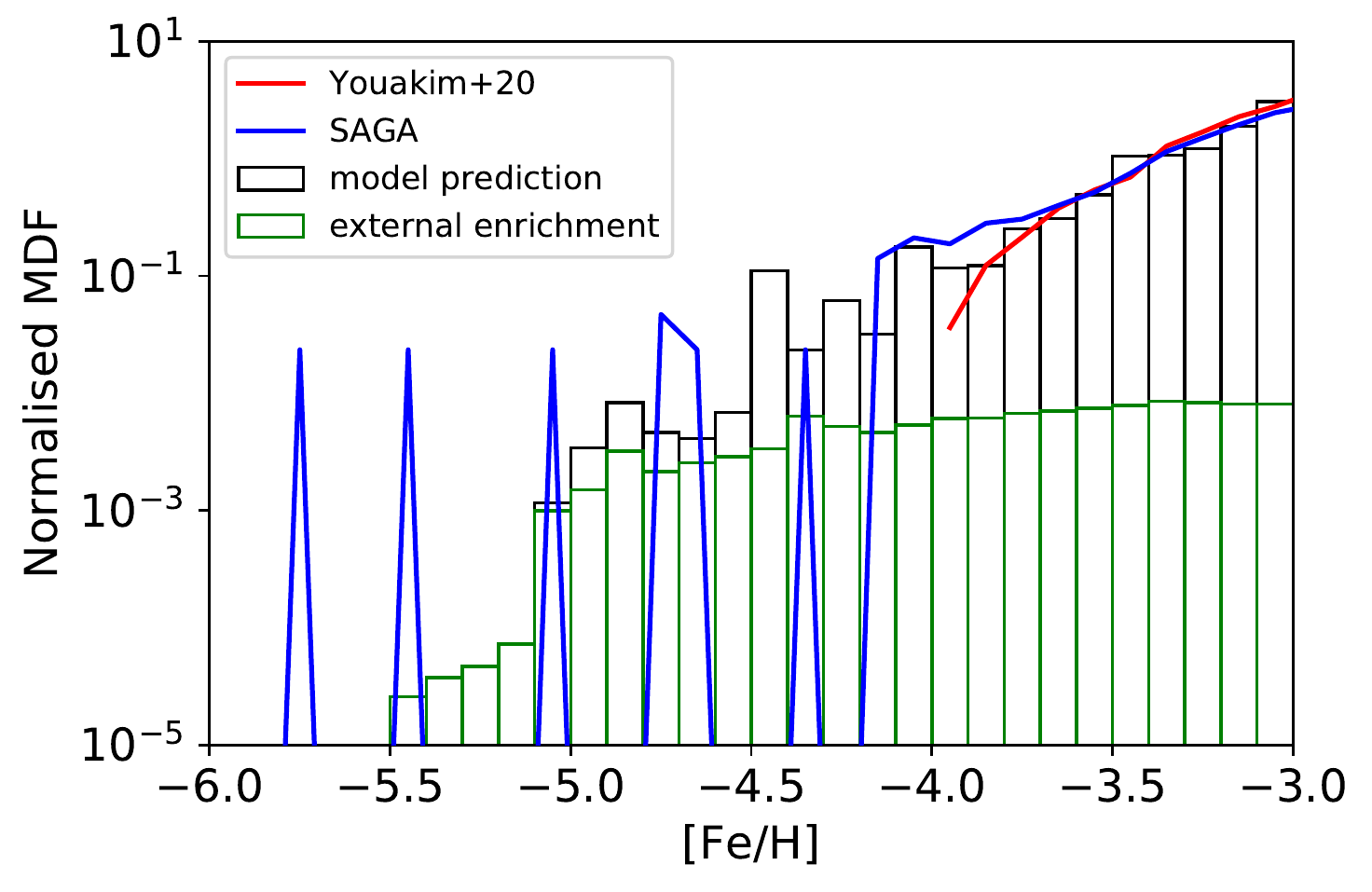}
        \caption{Averaged MDF of 30 merger trees with the fiducial model parameters. The black histogram is the model prediction. The green histogram is the metallicity distribution function only from external enrichment. The red curve is the de-biased MDF obtained by \citet{2020Youakim_MDF} and the blue curve is the MDF from SAGA database \citep{saga}.}
        \label{fig:MDF_example}
    \end{figure}
    
    In Fig.~\ref{fig:MDF_example} we compare the calibrated MDF and observed MDF. At [Fe/H]$ > -4.5$ range, internal enrichment is dominant. The metallicity inhomogeneity only plays a minor role on this metallicity range. In our calibration we only compare MDF at [Fe/H]$ > -4$ range. Therefore, the metallicity inhomogeneity only has a small effect on the Pop~III IMF. 
    
    \begin{figure}
	\includegraphics[width=\columnwidth]{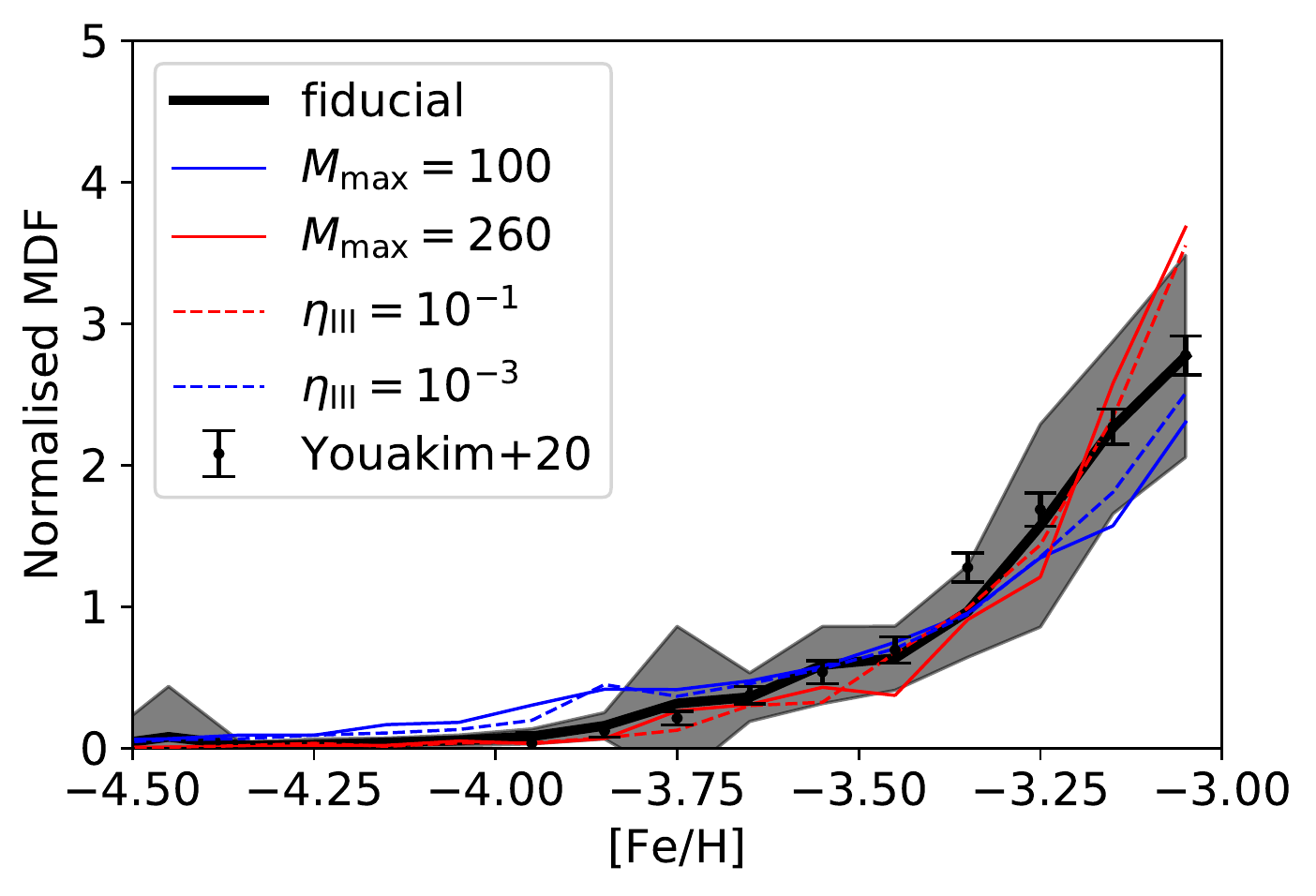}
    \caption{MDF comparison among various model parameters. The black dots represent the observed MDF \citep{2020Youakim_MDF}. To show the approximate error from sampling, we plot Poisson errors assuming that the error is obtained by the number of observed stars in each bin. The black curve represents the fiducial model MDF and the shaded region shows the one sigma scatter of the 30 trees. The other curves are model predictions with one parameter modified from its fiducial value. The modeled MDFs are normalised by the mass of stars at [Fe/H] $ < -3$. The observed MDF is normalised by the number of stars at [Fe/H] $ < -3$.
    We can see that $M_\mathrm{max} = 100\Msun$ and $\eta _{III} = 10^{-3}$ predict too many stars at [Fe/H]$\ \lesssim -4.0$. These models tend to produce less metals per Pop~III star formation event. Such small metal mass events contribute too much to the formation of stars at [Fe/H]$\ < -4.0$. On the other hand, $M_\mathrm{max} = 260\Msun$ and $\eta _{III} = 10^{-1}$ predict too many stars in [Fe/H]$\ \gtrsim -3.25$, although they are consistent to the fiducial MDF within the scatter of ``Milky Way-like merger trees''. To eliminate such parameter sets we need to resort to different information sources.}
    \label{fig:MDFParameterExploration}
    \end{figure}

    In Fig.~\ref{fig:MDFParameterExploration} we show how a different choice of model parameters affects the MDF.
    For some choices of parameters (($M_\mathrm{max} = 100$) model,\ $(\eta_{\rm III} = 10^{-3}$) model presented with blue curves), the models predict too many stars at [Fe/H] $\ \lesssim -4.0$. This is the consequence of the decrease in the amount of metals produced by Pop~III stars. Stars with lower mass convert pristine gas to metals less efficiently than higher-mass stars. Also, a lower Pop~III star formation efficiency decreases the overall metal production from Pop~III stars. The decrease in metal mass from Pop~III stars consequently decrease the metallicity of second-generation stars, therefore the number of stars 
    at [Fe/H] $\ < -4.0$ increases.
    For other choices of parameters (($M_\mathrm{max} = 260$) model,\ $(\eta_{\rm III} = 10^{-1}$) model, presented with red curves), the opposite trend is observed. These models produce too much metals from Pop~III stars, directly enriching the host galaxy to [Fe/H] $\gtrsim -3$. In these models the fraction of stars in [Fe/H] $\lesssim -3.5$ is less than the observed stellar metallicity distribution function, although the difference is typically smaller than the scatter among different merger trees. This comparison suggests that the lower limits of $M_\mathrm{max}$ and $\eta_\mathrm{III}$ can be constrained well by our method, and to constrain the upper limits of these parameters we need additional information such as the (non-) detection of stars with PISNe abundance pattern \citep[see e.g.][]{salvadori19}.

    \subsection{Additional Observables}
\begin{figure}
        \centering
        \includegraphics[width=\hsize]{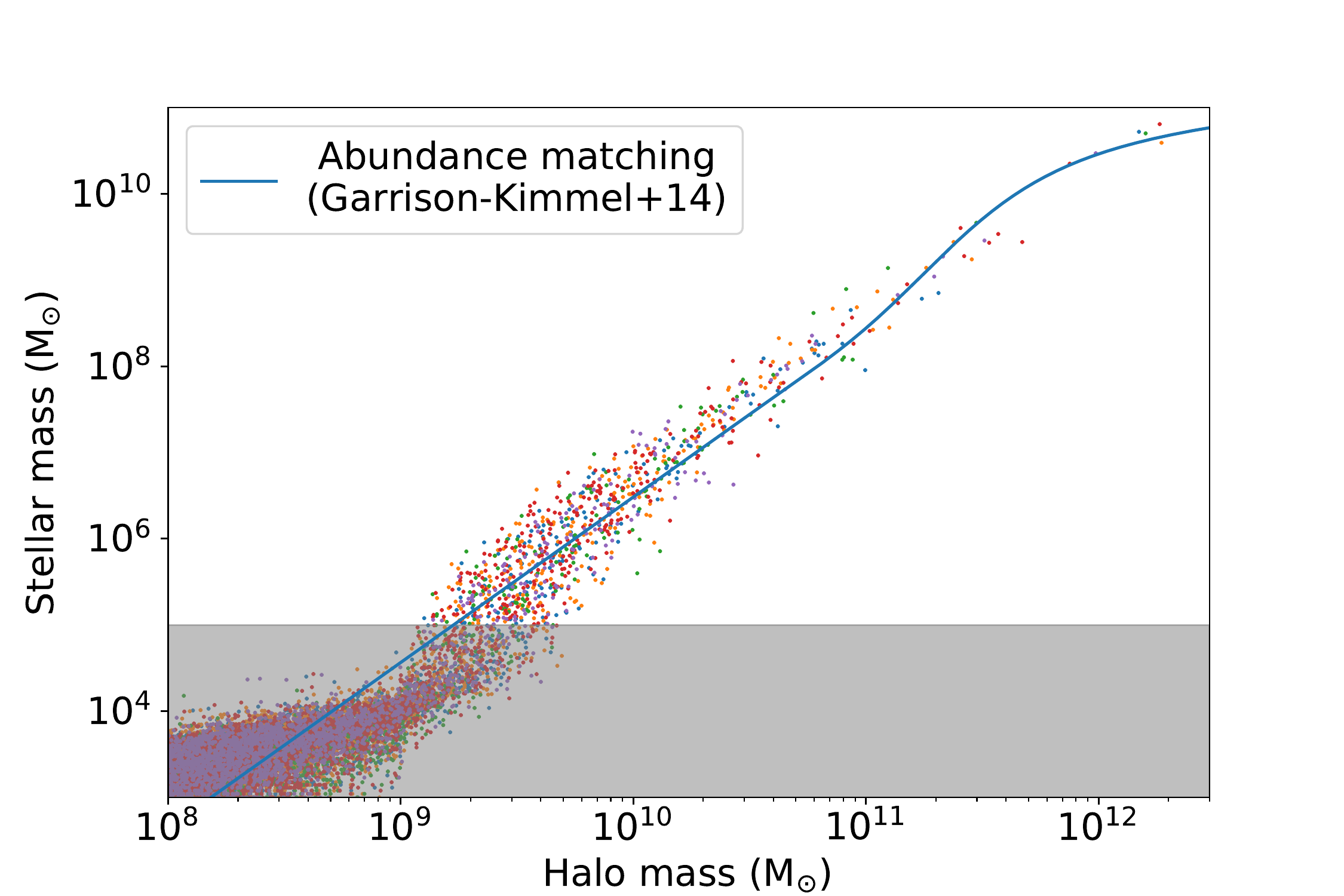}
        \caption{Comparison of stellar mass to halo mass. The dots are the simulated galaxies at $z=0$ in \textsc{a-sloth} with fiducial parameters. Five different colours correspond to five different merger trees, showing the tree-to-tree scatter. The solid line is the abundance matching relation from observations \citep{gk14} and the grey region indicates the stellar mass range in which the abundance matching prediction becomes unreliable.}
        \label{fig:SMHM}
    \end{figure}

    In Fig.~\ref{fig:SMHM} we show the stellar mass to halo mass relation at present day. The dots are the stellar mass at $z=0$ calculated with our fiducial model. The trees are sampled randomly. The solid line is the abundance matching relation derived by comparing the number of satellites around the MW with dark-matter-only simulations of MW-like haloes \citep{gk14}. In the shaded region (stellar mass $<10^{5} \Msun$), the abundance matching relation is not reliable due to poor sampling in the observation \citep{gk17}. Our fiducial model reproduce the stellar mass to halo mass relation reasonably well at $z=0$.
    
    Our model also reproduces the fraction of externally enriched halos as a function of redshift that is found in the Renaissance simulation: it increases from $\sim 10\%$ at $z=18$ to $~23\%$ at $z=12$. While this is not an observable, it is an additional independent crosscheck for our approach.
    
    \begin{figure}
        \centering
        \includegraphics[width=\hsize]{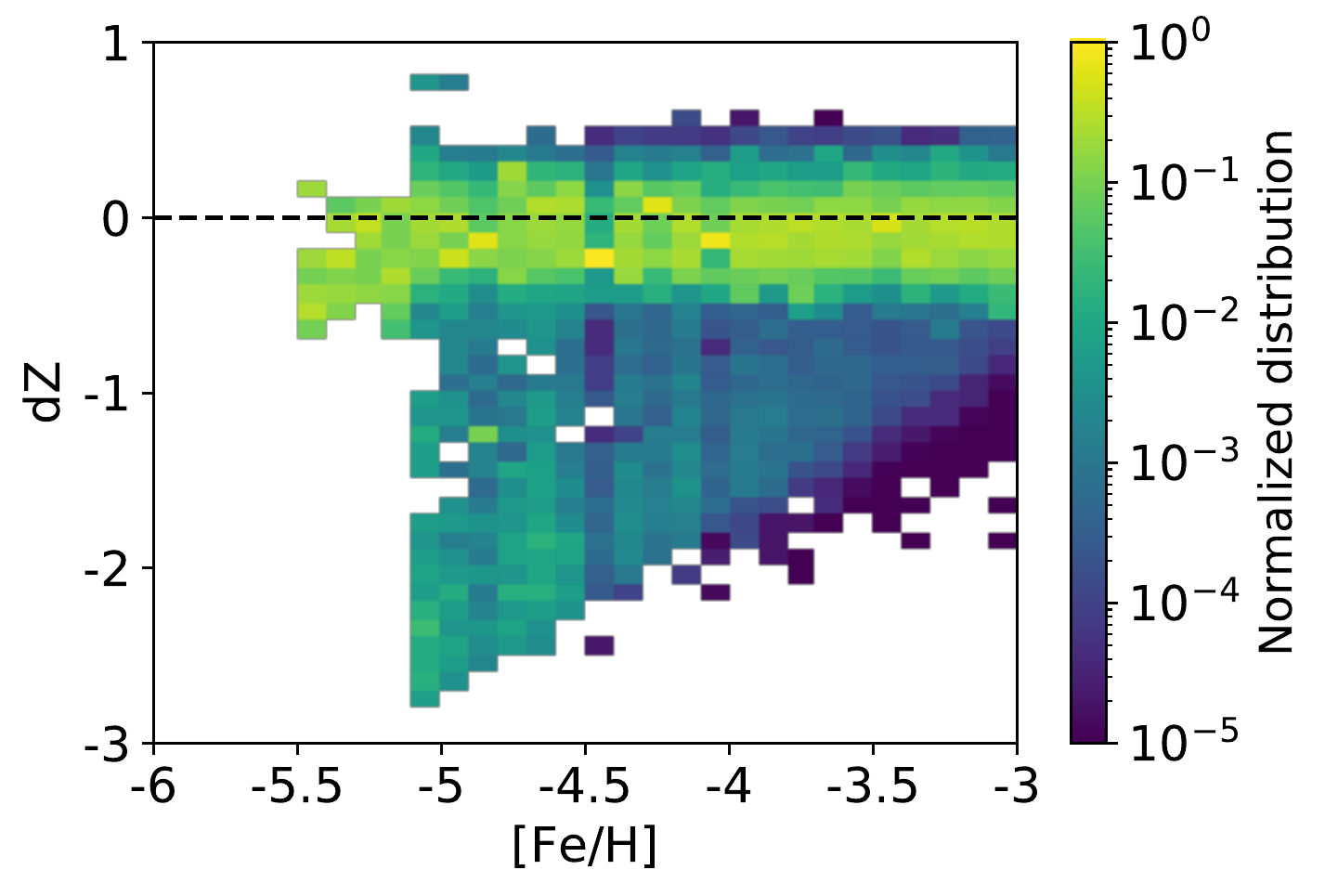}
        \includegraphics[width=\hsize]{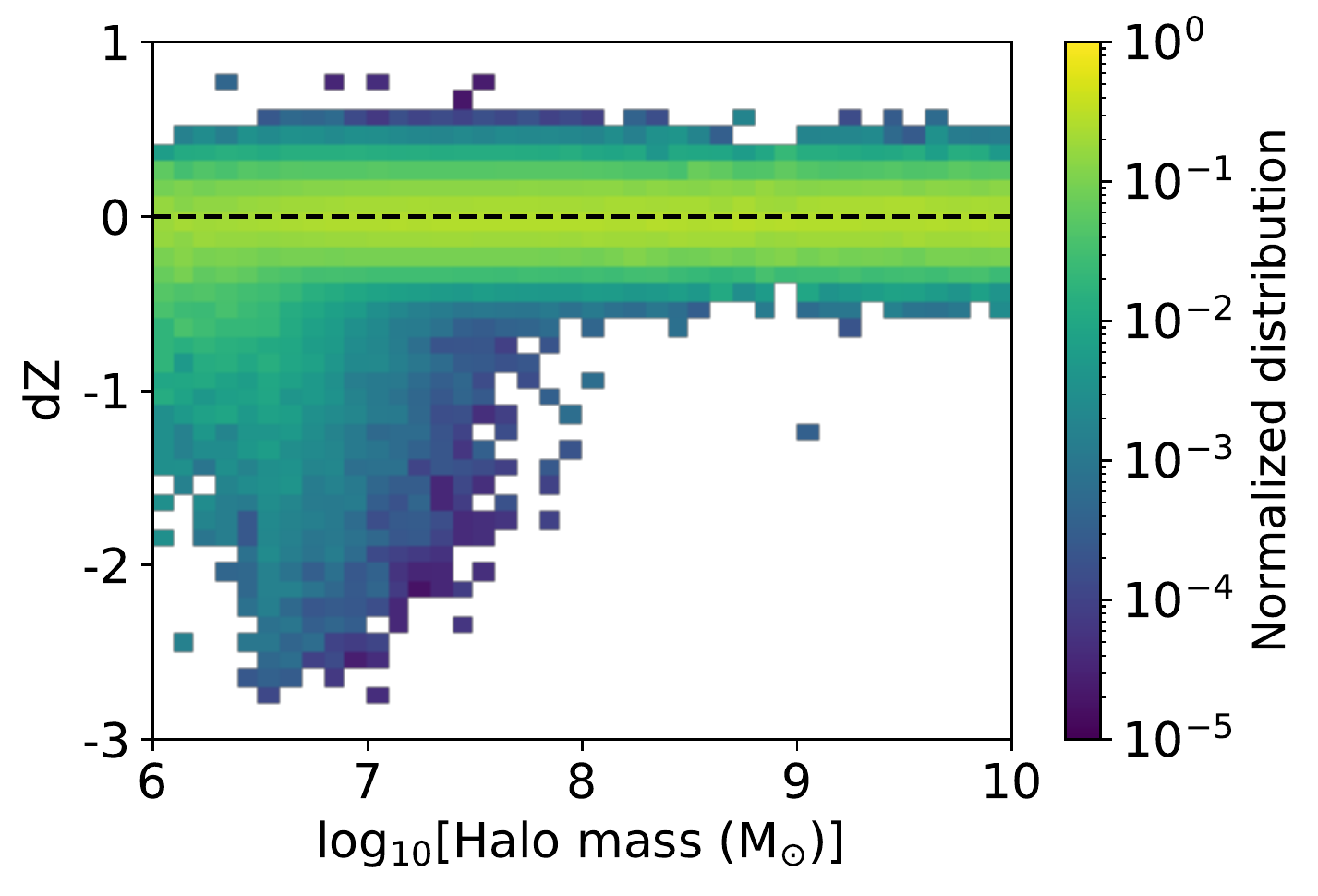}
        \includegraphics[width=\hsize]{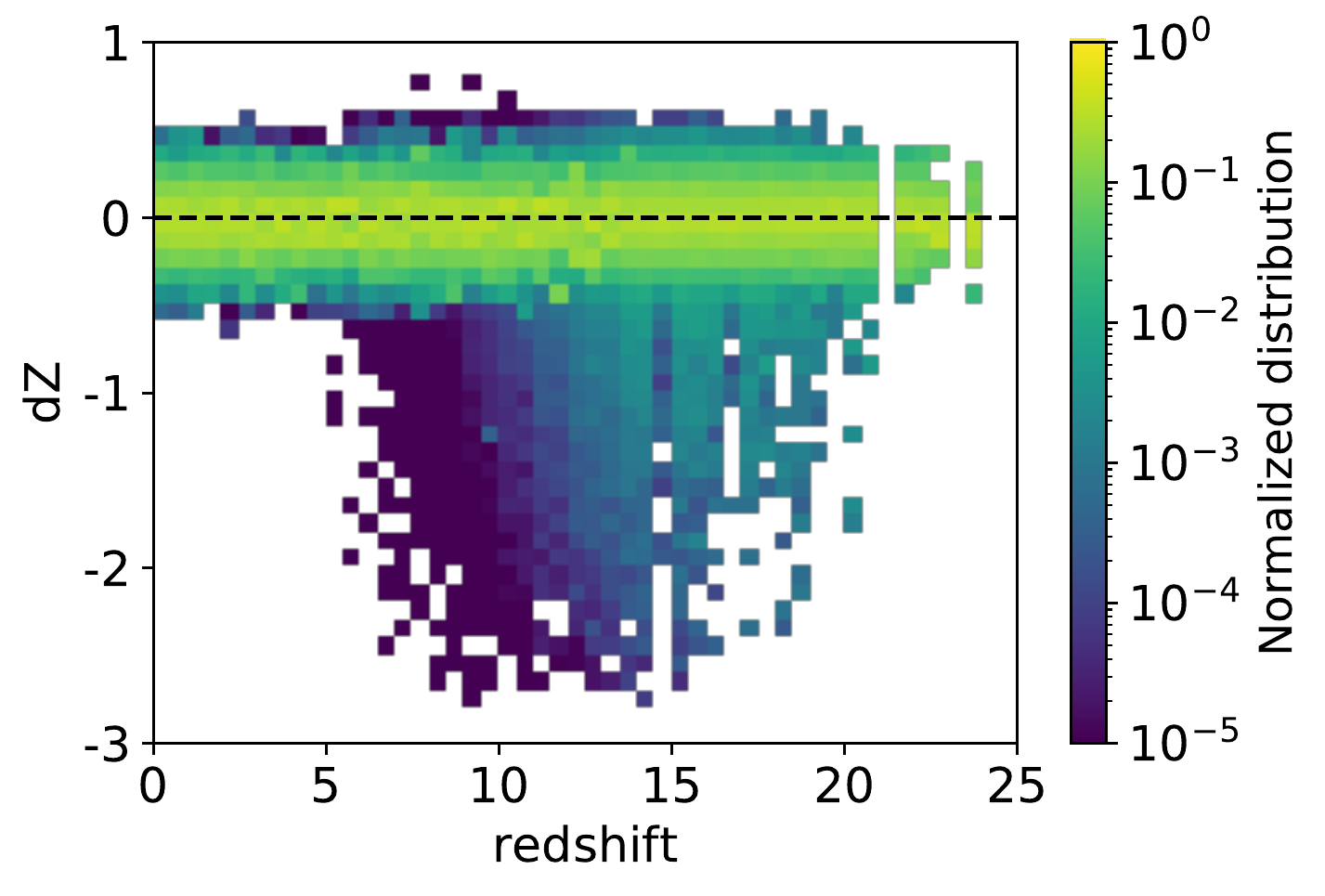}
        \caption{Distribution of $dZ$ as functions of various physical properties of stars. Top panel: for stars with [Fe/H] $< -4.5$, external enrichment is the dominant channel. Therefore, inhomogeneity is important for such ultra metal-poor stars. Middle panel: for large, matured halos ($M_\mathrm{halo} > 10^{8} \Msun$) star formation has already begun and therefore they are internally enriched. External enrichment is important when we consider star formation in small galaxies ($\sim 10^{7} \Msun$). Bottom panel: for star formation events at a very high redshift, external enrichment plays a role.}
        \label{fig:dZ_distributions}
    \end{figure}
    
    \subsection{When is $dZ$ important?}
    In Fig.~\ref{fig:dZ_distributions} we show the distribution of $dZ$ as functions of physical parameters. Since $dZ$ is most important in external enrichment, these panels indicate under what conditions external enrichment becomes important. For some stars with low [Fe/H] ($\lesssim -4$), $dZ$ is largely negative. This means that the metallicities of these stars would have been much higher if we assumed homogeneity. The clear cut at [Fe/H] $\sim -5.1$ comes from our Pop~II star formation criterion. For stars with [Fe/H] $ > -5.1$, we do not require carbon enhancement to form stars, whereas if [Fe/H] $ < -5.1$ we do require the carbon enhancement (see eq.~\ref{eq:transition_criterion}). Highly negative $dZ$ means high (naively calculated) [Fe/H], therefore less likely to be carbon enhanced and allowed to form stars. 
    
    The middle panel shows that in small halos $dZ$ can be largely negative. In massive galaxies ($\gtrsim 10^{8} \Msun$), star formation has already begun, therefore they are internally enriched and well mixed. Combining with our former analysis on the $dZ$ distribution, we find that the metallicity inhomogeneity is less significant in massive galaxies. 
    
    The bottom panel shows that $dZ$ is only important at high (although not the highest) redshifts. Low-redshift galaxies are homogeneous due to the same reason as the massive galaxies. The highest redshift galaxies ($z \gtrsim 20$) are homogeneous, because they do not have enough time after the first star formation events to experience external enrichment.
    In summary, inhomogeneous metal mixing is important in low-mass halos at high redshift that are about to form EMP stars.
    
\section{discussion}
We have performed the first cosmologically representative analysis of metal mixing in high redshift galaxies. We derived a physically motivated estimate of $dZ$, the metallicity difference between star-forming and all gas, and find that the distribution of $dZ$ is very different between haloes with stars and haloes without stars. This bimodality can be understood by assuming that two processes are at work: internal enrichment and external enrichment. Haloes without stars have not experienced any star-formation events, so they can be identified as externally enriched haloes. In external enrichment, the momentum of metal-rich wind is not strong. If a galaxy already has a dense gas cloud when the external enrichment takes place, it is expected that metals cannot penetrate into the dense gas. In such cases, $dZ$ is expected to be negative, indicating incomplete mixing. On the other hand, haloes with stars have experienced star formation at least once. The energy injection from SNe mix up the ISM in the host halo, which results in more homogeneous mixing between hydrogen and metals.

The formula we have obtained for $dZ$ suggests that the metallicity difference of star-forming gas cloud and average gas can be large in external enrichment (Eq. \ref{eq:Gaussian} and Eq. \ref{eq:PdZ_external}). We show that in this case 39 per cent of 
haloes have $dZ$ less than $-1$, which means more than 10 times metal-poorer than the average gas. A naive estimate of metallicity for stars formed in external enrichment can therefore overestimate the actual stellar metallicity. On the other hand, in internal enrichment, the average inhomogeneity is small: the fitted distribution of $dZ$ has a mean of $\mu = -0.03$ and a standard deviation of $\sigma=0.15$. 

The absence of any variable that has strong correlation to $dZ$ other than metallicity leads us to the conclusion that the metallicity difference between dense gas and average halo gas is intrinsically stochastic. One explanation is the missing stability of star formation and gas circulation.
The stochasticity of star formation in small haloes is pointed out by \citet{xu16} and \citet{sharma19}. Since small haloes have shallow gravitational potential wells, they easily lose their gas by stellar feedback. The stochasticity of metallicity difference between star-forming gas and average gas can be related to the stochasticity of star formation. Moreover, the first galaxies had not yet enough time to develop self-regulated star formation and therefore correlations between the involved physical quantities.

Despite the large metallicity difference in external enrichment, the predicted IMF is not sensitive to this difference. This can be understood because we compare the MDF mainly at $-4 < \mathrm{[Fe/H]} < -3$, where internal enrichment is the dominant channel and we mostly apply the corresponding distribution with a mean value of $\mu = -0.03$. Therefore, the statistical average over many haloes cancels out the inhomogeneity of each halo.

Taking inhomogeneous metal mixing into account does not have a significant influence on current observables. However, we show that the MDF at [Fe/H]$ < -4.5$ is affected by inhomogeneous metal mixing in the first galaxies (compare Fig.~\ref{fig:MDF_example}). Future observations of more ultra metal-poor stars can confirm or falsify our model of inhomogeneous metal mixing by discriminating MDFs at [Fe/H]$ < -4.5$.

Our derived IMF is in general agreement with the model by \citet{sarmento19} who show that the Pop~III IMF was dominated by stars in the mass range $20-120\Msun$, by comparing the radiative and chemical imprint of the first stars to observations. Our upper mass limit of $M_\mathrm{max} = 180\Msun$ is a bit higher than earlier estimates by \citet{deB17} who show that metal enrichment of EMP stars from PISNe is very rare, and by \citet{jeon17}, who simulate the chemical composition of MW satellites by adopting a Pop~III IMF up to $M_\mathrm{max}= 150\Msun$.

We also examined the fraction of carbon-enhanced metal-poor (CEMP) stars as a function of metallicity. CEMP stars are a very prominent sub-class of EMP stars with [C/Fe]$>0.7$ \citep{beers92,aoki07,lee13,salvadori15,sharma18}. The fraction of CEMP stars increases with decreasing metallicity \citep{yong13,placco14,yoon16}, which places them as prototypes of second-generation stars \citep{hansen16}.

While we reproduce the general trend of the CEMP-no (CEMP stars without enhancement of neutron-capture elements) fraction, we do not reproduce the fraction of CEMP-no stars in the metallicity range $-5\lesssim$[Fe/H]$\lesssim -3$ with our fiducial model. Our model predicts that only 0.4 per cent of stars with $-4.5 < \mathrm{[Fe/H]} < -3$ are CEMP-no stars, which is below the reported values by \citet{norris19} which derived that 12 per cent is truly CEMP-no stars even after 3-D NLTE corrections.

Recently, also \citet{komiya20} suggests that it is quite difficult to reproduce both the MDF and the fraction of CEMP-no stars if faint SNe are the only considered channel for the formation of CEMP-no stars. While we include faint SNe based on \citet{ishigaki14}, we miss, for example, mass transfer from a binary companion \citep{arentsen19}, carbon enrichment from rotating massive stars \citep{choplin19b}, differential mixing of carbon and iron \citep{frebel14,hartwig19}, or aspherical SN explosions \citep{tominaga07,ezzeddine19}. We have not yet included these additional channels because their nature and relative contribution is not well understood and a topic of ongoing research \citep{yoon19}. Therefore, we expect that our current model can only provide a lower limit to the CEMP fraction.

\subsection{Caveats}
    Hydrodynamical simulations are limited by numerical resolution. We confirmed that the resolution is sufficient to capture the metallicity structure of dense gas up to $100 \ccm$, because we see almost no difference among different choices of this density threshold in the range of [$1\ccm, 100\ccm$]. However, in order to follow metallicity difference between stars and gas completely, one should analyse the metallicity of denser gas, up to protostar formation. In our work, we could not follow the dense gas cloud phase, where stars are formed, due to limited resolution.

    Furthermore, \citet{schauer19b} have shown that around 1000 resolution elements per halo are required to properly resolve the onset of star formation. Thus, star formation in the Renaissance simulations is likely to be artificially delayed, and the mixing behaviour in the smallest haloes may not be captured in our model.
    
    The absolute and relative metal yields from Pop~III SNe are subject to uncertainties. The amount of ejected metals depends on, e.g., the explosion energy or rotation. Our model yields for Pop~III SNe are based on \citet{nomoto13}. The mass-dependent explosion energies for these SNe are calibrated based on observed explosion energies at higher metallicity. However, the explosion energies of Pop~III SNe, and therefore the effective metal yields, may not be a monotonic function of the progenitor mass, but rather a distribution of explosion energies \citep{ishigaki18}. Hence, while the IMF-averaged metals are more reliable, the metal yields for individual stars may differ from our implementation due to stochastic differences in the explosion energy or stellar rotation.

\section{Conclusions}

    In this work, we have investigated the effect of inhomogeneous metal mixing on the metallicities of EMP stars. The inhomogeneity of metallicity has not been well understood and is often ignored in semi-analytical approaches. We analyzed the cosmological hydrodynamic ``Renaissance Simulations'' \citep{Oshea15} to gain insight into the metallicity of star-forming gas in the first galaxies. The aim is to predict the stellar metallicity based on the metal mass and the gas mass in the halo, taking inhomogeneity into account.
    
    The analysis of hydrodynamical simulations shows that the metallicity of star-forming gas can be different from the average metallicity in a halo.
    Our analysis suggests that the difference of metallicity between dense gas and average gas, $dZ$, behaves systematically different for haloes with and without stars (Fig.~\ref{fig:dZ_Mstars_bimodality}). For starless haloes, $dZ$ is largely negative (typically about or more than 1\,dex), and it increases with overall metallicity, suggesting that it is difficult to enrich already dense gas clouds with external enrichment. For haloes with stars, $dZ$ tends to be close to zero \citep[see also][]{cote18}, however slightly negative with 0.15 dex scatter, which is comparable to observational uncertainties. We do not find other correlations to $dZ$, highlighting its stochastic nature. 
    The small metallicity difference $dZ$ in internal enrichment suggests that the homogeneous assumptions inside halo in many of existing SAMs are a good approximation. However, one should be cautious in externally enriched halos, where the difference between metallicity of all and of star-forming gas exceeds 1\,dex in 39 per cent of the cases.
    
    Taking metallicity inhomogeneity into account, we calibrated our SAM, \textsc{a-sloth}, and explored various sets of Pop~III IMF-generating parameters. The best IMF is a function with $dN/dM \propto M^{-0.5}$ between [2, 180] $\Msun$. The predicted IMF did not change significantly by taking inhomogeneity into account.
    The uncertainty and degeneracy in other parameters such as Pop~III star formation efficiency can change the prediction on Pop~III IMF \citep{cote17}. This degeneracy can be resolved if we can obtain an independent estimate on Pop~III star formation efficiency, such as direct observations of Pop~III-dominated galaxies at high redshift with next generation telescopes.

\section*{Acknowledgements}
Numerical computations were carried out on Cray XC50 at Center for Computational Astrophysics, National Astronomical Observatory of Japan. We thank the anonymous referee for helpful and constructive comments on this manuscript.
YT is grateful for the hospitality of the Department of Astrophysical Sciences at Princeton University. YT's visit was supported by the University of Tokyo-Princeton strategic partnership grant.
This work was supported by JSPS KAKENHI Grant Number 19K23437.
MM was supported by the Max-Planck-Gesellschaft via the fellowship of the International Max Planck Research School for Astronomy and Cosmic Physics at the University of Heidelberg (IMPRS-HD).

We are grateful to the Renaissance collaboration for kindly sharing their simulation and to the Caterpillar collaboration for providing their dark matter merger trees. We also thank Naoki Yoshida, John Wise, and Conor Omand for stimulating discussions and helpful comments on the paper draft and we thank Kris Youakim for sharing his MDF. The majority of the plots were done with \textsc{yt}. YT wishes to thank the \textsc{yt} community for their hard work, which facilitated this research. We acknowledge the work by Takuma Suda to maintain the SAGA database.

\bibliographystyle{aasjournal}
\bibliography{MetalMixing}

\end{document}